%% file: main.tex
\documentclass[12pt]{iopart}

\usepackage{subcaption}
\usepackage{wrapfig}
\usepackage{amssymb}
\usepackage{amsfonts}
\usepackage{latexsym}
\usepackage[linesnumbered,lined,boxed,commentsnumbered]{algorithm2e}
\usepackage{bm}	
\usepackage{graphicx} 
\usepackage{hyperref}
\usepackage{bigints}
\usepackage{lipsum} 
\usepackage{float}
\usepackage{booktabs}
\usepackage{siunitx}
\usepackage{outlines}

\SetAlFnt{\small}

\usepackage{url}
\usepackage{xcolor}
\definecolor{newcolor}{rgb}{.8,.349,.1}

\usepackage{listings}

\providecommand{\keywords}[1]{\textbf{\textit{keywords---}} #1}

\definecolor{mGreen}{rgb}{0,0.6,0}
\definecolor{mGray}{rgb}{0.5,0.5,0.5}
\definecolor{mPurple}{rgb}{0.58,0,0.82}
\definecolor{backgroundColour}{rgb}{0.95,0.95,0.92}

\lstdefinestyle{CStyle}{
    backgroundcolor=\color{backgroundColour},   
    commentstyle=\color{mGreen},
    keywordstyle=\color{magenta},
    numberstyle=\tiny\color{mGray},
    stringstyle=\color{mPurple},
    basicstyle=\footnotesize,
    breakatwhitespace=false,         
    breaklines=true,                 
    captionpos=b,                    
    keepspaces=true,                 
    numbers=left,                    
    numbersep=5pt,                  
    showspaces=false,                
    showstringspaces=false,
    showtabs=false,                  
    tabsize=2,
    language=C
}

\begin{document}

\title[Novatron: Theory and Simulation]{Novatron: Equilibrium and Stability}

\author{K Lindvall$^*$, R Holmberg, K Bendtz, J Scheffel, J Lundberg}
\address{Novatron Fusion Group AB}
\ead{$^*$kristoffer.lindvall@novatronfusion.com}

\begin{abstract}
The Novatron is a fusion concept characterized by its axisymmetric mirror-cusp magnetic topology. The magnetic field exhibits good curvature and a high mirror ratio. Plasma equilibrium profiles for the Novatron are obtained by solving an axisymmetric guiding-center anisotropic boundary-value problem. These profiles are then analyzed with respect to several MHD stability criteria, including the mirror, firehose, and interchange conditions. A generalized Rosenbluth and Longmire MHD interchange criterion, where anisotropic pressure variations along flux tubes are allowed for, is subsequently employed for determining stable MHD equilibria. Additionally, a corresponding CGL double adiabatic interchange criterion is investigated for obtaining stable equilibria in the collisionless limit, both theoretically and numerically using the large scale Hybrid Particle-In-Cell code WarpX.
\end{abstract}


\section{Introduction}
The development of magnetically confined fusion devices presents challenges in creating a viable reactor concept. A main challenge, for both closed-field line devices, like  Tokamaks and Stellarators, and open-field line devices such as magnetic mirrors, is that of plasma stability \cite{freidberg_ideal_2014, helander_stellarator_2012}. The stability often depends critically on the curvature of the magnetic field lines and their relation to the pressure gradient. If the magnetic field lines have an unfavorable curvature (concave toward the high-pressure region), pressure-driven instabilities - most notably interchange and ballooning modes - can arise, degrading the performance of the device.

\begin{figure}[H]
    \centering
    \includegraphics[width=0.7\textwidth]{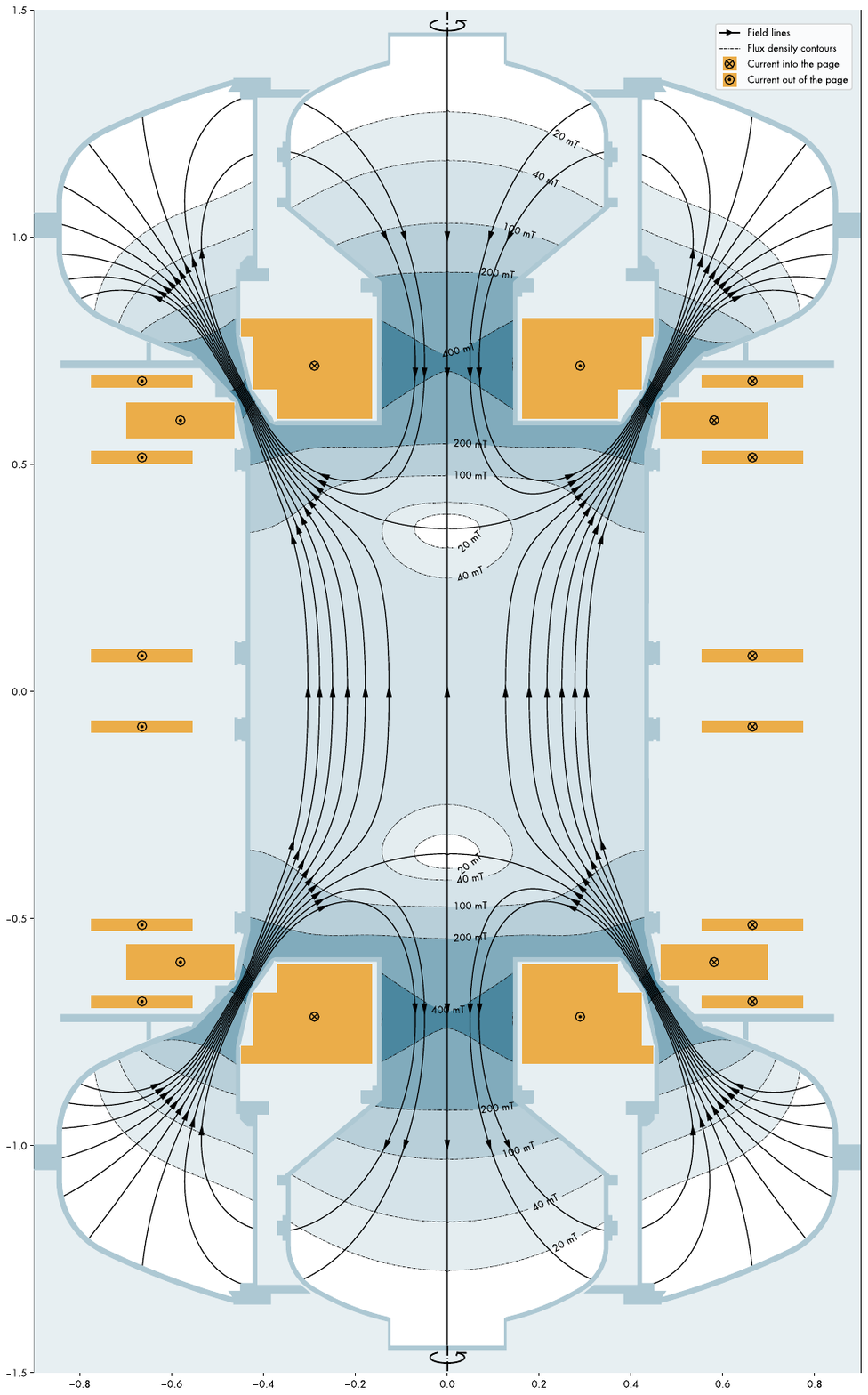}
    \caption{A base-line magnetic field configuration of the first generation Novatron device, N1.}
    \label{fig:N1MagneticField}
\end{figure}

The Novatron is a fusion device that utilizes open-field line confinement and thus builds on the established principles of magnetic mirror confinement. Its magnetic topology extends the double-cusp configuration \cite{hamamatsu_numerical_1983}, making it a hybrid between a classical magnetic mirror and a bi-conic cusp. A representation of the Novatron magnetic field topology can be seen in Figure \ref{fig:N1MagneticField}. As a hybrid the Novatron seeks to combine the axial particle confinement times of a classical mirror with the MHD stability of a cusp. To further increase the confinement time the Novatron can be equipped with tandem-cells for electrostatic ion confinement, along with ponderomotive plugging between the cells. 

A first generation Novatron device, as seen in Fig. \ref{fig:N1MagneticField}, has recently been commissioned by NFG Novatron Fusion Group at KTH Royal Institute of Technology, Stockholm, Sweden. For a detailed explanation of the Novatron concept and a comparison to other magnetic confinement devices, see Ref. \cite{jaderberg_introducing_2023}. The properties of the Novatron, summarized in this publication, include anisotropic MHD equilibrium stability, operation at high-$\beta$ values, minimal neoclassical transport, and high mirror-ratios for improved axial confinement. 

In 1964, Northrup and Whiteman \cite{northrop_minimum-_1964} derived a set of partial differential equations that model axisymmetric anisotropic equilibrium pressure profiles. These equilibrium equations were solved by Killeen and Whiteman in 1966 \cite{killeen_computation_1966} for pressure profiles dependent on the total magnetic field, that is $p=p(B)$, as described by Taylor in 1963 \cite{taylor_stable_1963}. The guiding center theory behind the equilibrium equations was then expanded by Grad in 1967 \cite{grad_guiding_1967}, allowing for the solution of pressure profiles dependent on both magnetic field $B$ and stream function $\psi$. These equations were solved by Fischer and Killeen in 1971 \cite{fisher_finite_1971} for toroidal anisotropic plasma configurations. Here we use a similar numerical approach as that of Fischer and Killeen \cite{fisher_finite_1971} and solve the equilibrium equations for an open-field Novatron plasma configuration.

With the proper anisotropic equilibrium pressure profiles in place, a stability analysis can be performed. This analysis builds upon the theoretical work on the interchange instability of magnetic mirrors done by Rosenbluth and Longmire in 1957 \cite{rosenbluth_stability_1957}, wherein a line integral interchange criterion is derived by assuming a specific and restrictive instability where two flux tubes interchange. As noted by Hastie \cite{hastie_interchange_1962}, the interchange criterion derived in Ref. \cite{rosenbluth_stability_1957} is of first order and applicable on the outer edge of a plasma. Since the Novatron is a variation of a double-cusp, it features two regions of magnetic null, and hence the magnetic field is of the well-type. If the plasma pressure decreases towards the symmetry axis, a similar interchange phenomenon can occur on the inside as that on the outside of a classical mirror. 

To address the specific magnetic geometry in a Novatron we have derived two second order interchange criteria. The first is a generalized version of the Rosenbluth and Longmire criteria, where we now allow the pressures to vary along the flux tubes whilst keeping higher-order terms. Furthermore, the MHD criteria derived here can be applied globally. The second criteria is derived in the same manner employing the Chew-Goldberger-Low (CGL) double adiabatic equations, which assumes a collisionless plasma.

The paper is organized as follows: Section 2 presents the mirror equilibrium partial differential equations, to be solved for anisotropic pressure profiles. The equilibrium profiles are then analyzed for their critical $\beta_c$ values and diamagnetic effects. Section 3 presents the derivation of two new interchange criteria, which are applied to the anisotropic equilibria. Section 4 is devoted to high-$\beta$ hybrid Particle-in-Cell simulations in the magnetic field configuration of the Novatron 1. Here, the progress and development of the hybrid-PIC code WarpX is presented, including the inclusion of static external magnetic fields and complex polygon embedded walls to simulate vacuum vessels and limiters. This is followed by a discussion in Section 5 and conclusion in Section 6.

\section{Macroscopic Mirror Equilibrium}
The primary mathematical model for anisotropic plasma equilibria is derived from the magnetohydrodynamic (MHD) equations, incorporating an anisotropic pressure tensor. These equations are expressed as follows, employing dyadic notation:

\begin{align}
    \mathbf{J}\times\mathbf{B} = \nabla \cdot \mathbf{P}
\end{align}
\begin{align}
    \nabla \times \mathbf{B} = \mu_0\mathbf{J}
\end{align}
\begin{align}
    \nabla \cdot \mathbf{B} = 0
\end{align}
\begin{align}
    \mathbf{P} = p_{\perp}\mathbf{I} + \frac{1}{B^2}(p_{\parallel} - p_{\perp})\mathbf{BB}
\end{align}
where $\mathbf{B}$ is the magnetic field, $\mathbf{J}$ is the current density, $\mathbf{P}$ is the anisotropic pressure tensor consisting of the perpendicular and parallel pressures, $p_{\perp}$ and $p_{\parallel}$, respectively, the vacuum permeability $\mu_0$, and $\mathbf{I}$ is the identity dyadic tensor $\mathbf{I} \equiv \hat{\mathbf{e}}_1\hat{\mathbf{e}}_1 + \hat{\mathbf{e}}_2\hat{\mathbf{e}}_2+ \hat{\mathbf{e}}_3\hat{\mathbf{e}}_3$.

The MHD equilibrium equations have earlier been applied to mirror-cusp configurations (where there is no magnetic null) \cite{taylor_stable_1963, taylor_equilibrium_1964}.
In an axially symmetric system the above equations can be combined \cite{grad_guiding_1967, grad_toroidal_1967} as
\begin{align} \label{eq:pde_mirror_equil}
    -\frac{\partial^2 \psi}{\partial z^2} - r \frac{\partial}{\partial r}\Big(\frac{1}{r}\frac{\partial \psi}{\partial r}\Big) = \frac{1}{\sigma^2}g'(\psi) + \frac{r^2}{\sigma}\frac{\partial p_{\parallel}}{\partial \psi} + \frac{1}{\sigma}\Big[\frac{\partial \sigma}{\partial z}\frac{\partial \psi}{\partial z} + \frac{\partial \sigma}{\partial r}\frac{\partial \psi}{\partial r}\Big]
\end{align}
where $g(\psi) = (1/2)\sigma^2r^2B_{\theta}^2$ and $\sigma = 1/\mu_0 + B^{-2}(p_{\perp} - p_{\parallel})$. Here the magnetic field and stream functions are defined as the sum of the vacuum field and plasma induced field, i.e., $\mathbf{B} = \mathbf{B}_p + \mathbf{B}_v$ and $\psi = \psi_p + \psi_v$. Introducing the external fields into Eq. \eqref{eq:pde_mirror_equil} gives

\begin{align} \label{eq:2}
    -\frac{\partial^2 \psi_p}{\partial z^2} - r \frac{\partial}{\partial r}\Big(\frac{1}{r}\frac{\partial \psi_p}{\partial r}\Big) - \frac{1}{\sigma}\Big[\frac{\partial \sigma}{\partial z}\frac{\partial \psi_p}{\partial z} + \frac{\partial \sigma}{\partial r}\frac{\partial \psi_p}{\partial r}\Big] = \frac{1}{\sigma^2}g'(\psi) + \frac{r^2}{\sigma}\frac{\partial p_{\parallel}}{\partial \psi} + \frac{r}{\sigma}\Big[B_{zv}\frac{\partial \sigma}{\partial z} - B_{rv}\frac{\partial \sigma}{\partial r}\Big].
\end{align}

\subsection{Novatron Equilibrium}
We wish to solve Eq. \eqref{eq:2} in order to determine realistic plasma pressure profiles given an external vacuum magnetic field. Ideally, one would solve for self-consistent equilibria using a transport code that would narrow down the possible families of equilibria. However, to simplify the analysis, we can make certain assumptions about the variable dependence to obtain analytical expressions for the pressure profiles.

We proceed by assigning a form of the parallel pressure profiles
\begin{align}
    p_{\parallel}(\psi, B) & = A(\psi) p_{\parallel}(B).
\end{align}
We will discuss three models that satisfy the equilibrium Eq. \eqref{eq:pde_mirror_equil}. These are:

\noindent Taylor's model \cite{taylor_stable_1963, taylor_special_2015},
\begin{align}
    p_{\parallel}(B) & = CB(B_1 - B)^M,        \\
    p_{\perp}(B)     & = CMB^2(B_1 - B)^{M-1}, ~~~ M\geq2
    \label{eq:taylor}
\end{align}
Rensink's model \cite{anderson_calculations_1982,anderson_self-consistent_1984},
\begin{align}
    p_{\parallel}(B) & = CB^{-M}(B_1-B)^{M+3/2},        \\
    p_{\perp}(B)     & = CB^{-M}(B_1-B)^{M+1/2}(B/2 + (M+1)B_1),  ~~~ M\geq1/2
    \label{eq:rensink}
\end{align}
Cutler's model \cite{anderson_calculations_1982},
\begin{align}
    p_{\parallel}(B) & = C(1 - (B/B_1) + (B/B_1)\text{ln}(B/B_1)),        \\
    p_{\perp}(B) & = C(1-(B/B_1)),
    \label{eq:cutler}
\end{align}
with all models above setting $p_{\parallel} = p_{\perp} = 0$ for $B > B_1$, where $B_1$ is a maximum magnetic field strength cut-off.

\begin{figure}[H]
    \centering
    \includegraphics[width=0.6\textwidth]{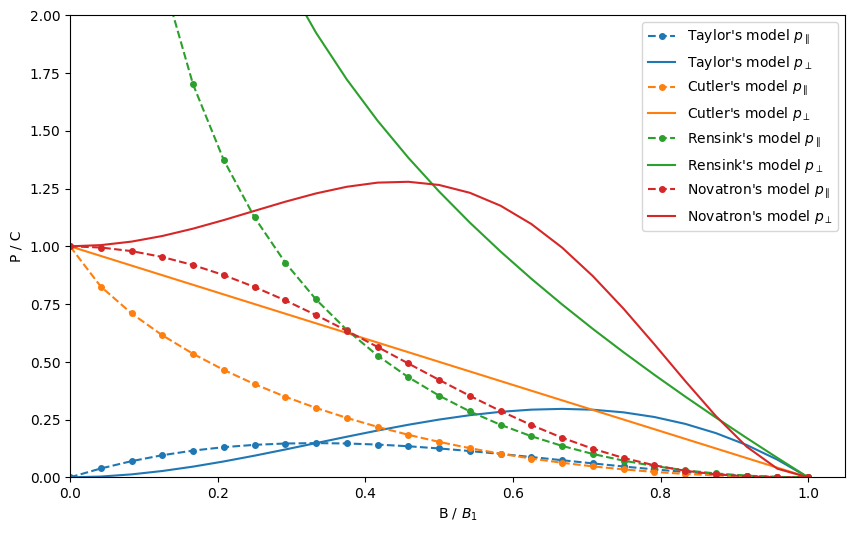}
    \caption{Comparison of the magnetic field strength dependence of the anisotropic pressure profiles: $M=2$ Taylor's model Eq. \eqref{eq:taylor}, $M=0.5$ Rensink's model Eq. \eqref{eq:rensink}, -- Cutler's model Eq. \eqref{eq:cutler}, and $M=3$ Novatron's model Eq. \eqref{eq:Novatron}.}
    \label{fig:PressureModels}
\end{figure}

A comparison of these models can be found in Fig. \ref{fig:PressureModels}. Taylor's model features a parallel pressure that has a maximum at a non-zero magnetic field value. The pressure profile also goes to zero as the magnetic field goes to zero. Although it is possible that a lower pressure develops in the magnetic null regions, it is unlikely that the pressure will be zero since plasma tends to drift towards lower magnetic field regions. The Rensink and Cutler models are analytical fits to transport code results of a classical mirror machine, and thus feature a more plausible pressure profile for low magnetic field values compared to Taylor's model Eq. \eqref{eq:taylor}. Rensink's profiles in this case diverge in the magnetic-null region, and thus the Cutler profiles are an improvement in this regard, since they set a maximum parallel pressure cap at $B=0$. The model also features, as does Rensink's model, an increasing parallel pressure profile towards $B=0$. This could fit well with bi-conic cusps that feature maximum pressures in the center of the device. The Novatron, however, will benefit from a smooth tapering transition in the adiabatic regions further away from the axis, and thus a parallel pressure gradient $dp_{\parallel}/dB|_{B=0}=0$ will help facilitate that.

In order to construct new pressure profiles, an arbitrary guiding center distribution function $f(\epsilon,\mu)$, where $\epsilon$ and $\mu$ are the particle energy and adiabatic moments respectively, can be chosen that satisfies stability and equilibrium conditions. The distribution function can be chosen as an analytical fit from transport codes or experimental data. The pressure profiles can then be obtained by integrating the distribution function. 

Another strategy is to choose the parallel pressure profile and then use the guiding center relation derived by Grad \cite{grad_guiding_1967},
\begin{align}
    p_{\perp} = p_{\parallel} - B\frac{d p_{\parallel}}{d B},
    \label{eq:GC}
\end{align}
which directly determines the perpendicular pressure. Using Eq. \eqref{eq:GC}, a pressure profile for the Novatron configuration can be obtained that takes into account the magnetic-null regions and the transition between adiabatic and non-adiabatic regions,
\begin{align}
    p_{\parallel}(B) & = C(1 - (B/B_1)^2)^M,        \\
    p_{\perp}(B) &= C(1 - (B/B_1)^2)^{M-1}(1 + (2M - 1)(B/B_1)^2), ~~~~ M\geq2,
    \label{eq:Novatron}
\end{align}
for $B \leq B_1$, else $p_{\parallel} = p_{\perp} = 0$. This pressure profile ensures that $p_{\perp}(0)=p_{\parallel}(0) = C$.

The $\psi$-dependence can be chosen so as to limit the finite pressure profiles to pre-defined flux lines. One example of $A(\psi)$ is presented in \cite{fisher_finite_1971, killeen_computational_1986},
\begin{align}
    A(\psi) = \Big(\psi- \psi_0\Big)^N\Big(\psi_1 - \psi\Big)^N,
\end{align}
if $\psi_0 \leq \psi \leq \psi_1$, else $A(\psi) = 0$. The arbitrary constant $N$ is used to vary the steepness the plasma edge. Here we use a normalized form,
\begin{align}
    A_n(\psi) = \psi_n A(\psi) = 4^N\Bigg(\frac{\psi - \psi_0}{\psi_1 - \psi_0}\Bigg)^N\Bigg(\frac{\psi_1 - \psi}{\psi_1 - \psi_0}\Bigg)^N.
    \label{eq:norm_fisher_killen}
\end{align}
An alternative expression for a hollow profile is,
\begin{align}
    A_{hollow}(\psi) = \psi^2A(\psi).
\end{align}
Another interesting form is the flat profile with steep edges,
\begin{align}
    A_{flat}(\psi) = 
    \begin{cases}
        1, & \text{if } \psi\leq \psi_{flat} \\
        0.5\Bigg(1 + \text{cos}\Big(\frac{\pi(\psi - \psi_{flat})}{\psi_1 - \psi_{flat})
        }\Big)\Bigg),              & \text{if } \psi_{flat} < \psi \leq \psi_1 \\
        0, & \text{otherwise}
    \end{cases}
    \label{eq:flat}
\end{align}
Eq. \eqref{eq:flat} can be specified for both positive and negative $\psi$ values so that different edge depths can be chosen. Thus far, these flux-line limiting profiles $A(\psi)$ are heuristic. The full anisotropic pressure profiles presented here allow us to benchmark a vast span of promising equilibria that are of interest in a Novatron. It is then an engineering challenge to devise a fueling, ionization, and heating strategy that can reproduce the pressure profiles in an actual machine.

\begin{figure}[H]
    \centering
    \includegraphics[width=1.0\textwidth]{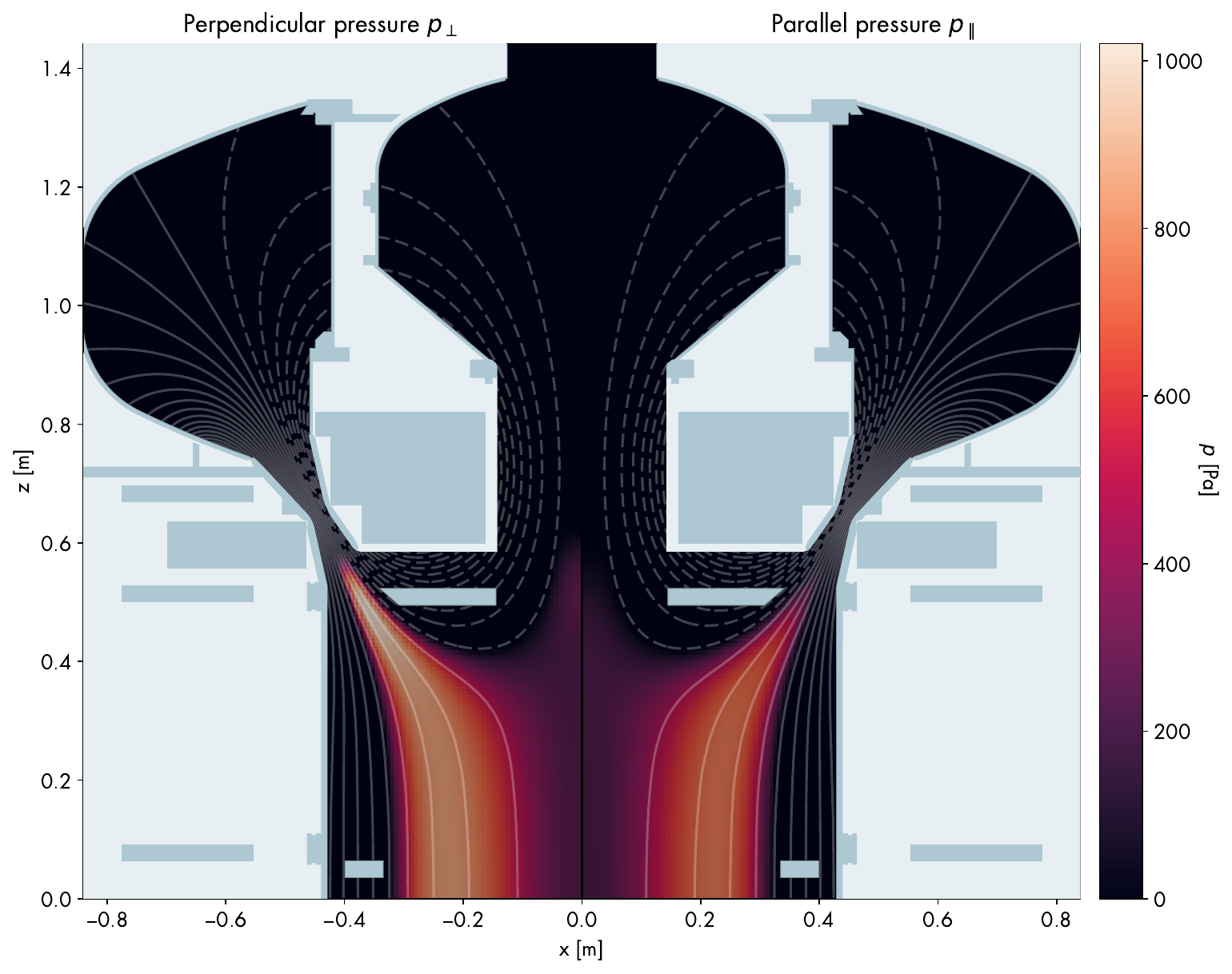}
    \caption{Perpendicular (left) and parallel (right) equilibrium pressure profile using Novatron pressure model Eq. \eqref{eq:Novatron}. with $B_1 = 0.34$, $\psi_0 = -0.00023$, $\psi_1 = 0.0028$, and $C = 800$, $M=3$, $N=1$, $n_z=214$, and $n_r=125$.}
    \label{fig:FullAnisotropicPressure}
\end{figure}

For numerical determination of equilibria, specifically the unknown plasma induced field $\psi_p$, Eq. \eqref{eq:pde_mirror_equil} is discretized with a standard finite difference method on a computational domain $\Omega = [0, L_r]\times[0, L_z]$ with $n_r$ and $n_z$ number of cells in each respective dimension. The boundary conditions can be chosen as Dirichlet, Neumann, or periodic. In Figure \ref{fig:FullAnisotropicPressure} a solution to Eq. \eqref{eq:pde_mirror_equil} can be seen where Dirichlet boundary conditions $\psi_p = 0$ are set on all boundaries except at $z=0$, where a Neumann condition is used to model axial symmetry.

We define the local $\beta$, as in previous work Ref. \cite{anderson_computation_1972},
\begin{align}
    \beta = 2p_{\perp} / (B_v^2 / \mu_0 + 2 p_{\perp}),
    \label{eq:local_beta}
\end{align}
that may approach unity. The local $\beta$ is then accompanied with subscripts such as $\beta_m$, which is the maximum local $\beta$ at the mid-plane ($z=0$). As an example in Figure \ref{fig:FullAnisotropicPressure}, with the equilibrium parameters $B_1 = 0.34$, $\psi_0 = -0.00023$, $\psi_1 = 0.0028$, and $C = 800$, $\beta_m$ is 35\%. 

We also defined the critical $\beta\equiv\beta_c$, which is the maximum $\beta_m$ that can be achieved with a specific equilibrium pressure profile. Since $\beta_c$ cannot be known a-priori, that is the induced magnetic field must be numerically computed, the macroscopic equilibrium Eq. \eqref{eq:2} must be recomputed with increasing $C$ values until the root solver no longer converges \cite{fisher_finite_1971}. When the highest $C$ value is reached for a given pressure profile then $\beta_c$ is computed, i.e. the maximum local $\beta$ at the mid-plane for the maximum $C$ value.

The $\beta_c$ values for pressure profiles Cutler Eq. \eqref{eq:cutler} and Novatron Eq. \eqref{eq:Novatron}, with $A_{n}(\psi)$ can be seen in Table \ref{table:An}, and with $A_{flat}(\psi)$ in Table \ref{table:Af}. We see similar $\beta_c$ values for all combinations. With the Cutler Eq. \eqref{eq:cutler} pressure profile $\beta_c$ is 62\%, and with the Novatron profile 63\% can be reached. Again, it should be noted, that $\beta_c$ is specific to the guiding center pressure profiles, and not the external magnetic configuration. It could be, for example, that a pressure profile exists which allows for a larger $\beta_c$. 

As with all magnetic confinement devices, the largest possible $\beta$ value must be assessed with the stability of that specific plasma profile. As an example, a high $\beta$ plasma profile can be theoretically obtained in a Tokamak, however, it would prove to be operationally unfeasible due to stability constraints.

\begin{table}[h!]
\centering
    \begin{tabular}{SSSSSS} \toprule
        {$M$} & {$N$} & {$\beta^{40,68}_c$} & {$\beta^{50,85}_c$} & {$\beta^{60,102}_c$} & {$P_{\parallel}$} \\ \midrule
        {-}  & 1 & 0.34 & 0.35 & 0.35 & {$C(1 - (B/B_1) + (B/B_1)\text{ln}(B/B_1))$} \\
        {-}  & 2 & 0.51 & 0.53 & 0.54 & {$C(1 - (B/B_1) + (B/B_1)\text{ln}(B/B_1))$} \\ \midrule
        {2} & 1 & 0.37 & 0.39 & 0.41 & {$C(1 - (B/B_1)^2)^M$} \\
        {2} & 2 & 0.56 & 0.53 & 0.53 & {$C(1 - (B/B_1)^2)^M$} \\
        {4} & 1 & 0.35 & 0.35 & 0.37 & {$C(1 - (B/B_1)^2)^M$} \\
        {4} & 2 & 0.55 & 0.59 & 0.57 & {$C(1 - (B/B_1)^2)^M$} \\ \bottomrule
    \end{tabular}
    \caption{Novatron $\beta_c^{n_r,n_z}$ values for $A_{n}(\psi)$.}
    \label{table:An}
\end{table}

\begin{table}[h!]
\centering
    \begin{tabular}{SSSSSS} \toprule
        {$M$} & {$\beta^{40,68}_c$} & {$\beta^{50,85}_c$} & {$\beta^{60,102}_c$} & {$\beta^{80,137}_c$} & {$P_{\parallel}$} \\ \midrule
        {-}  & 0.51 & 0.49 & 0.55 & 0.62 & {$C(1 - (B/B_1) + (B/B_1)\text{ln}(B/B_1))$} \\ \midrule
        {2} & 0.53 & 0.52 & 0.50 & 0.55 & {$C(1 - (B/B_1)^2)^M$} \\
        {4} & 0.52 & 0.51 & 0.57 & 0.63 & {$C(1 - (B/B_1)^2)^M$} \\
        {6} & 0.51 & 0.57 & 0.49 & 0.61 & {$C(1 - (B/B_1)^2)^M$} \\ \bottomrule
    \end{tabular}
    \caption{Novatron $\beta_c^{n_r,n_z}$ values for $A_{flat}(\psi)$.}
    \label{table:Af}
\end{table}

\begin{figure}[H]
    \centering
    \includegraphics[width=1.0\textwidth]{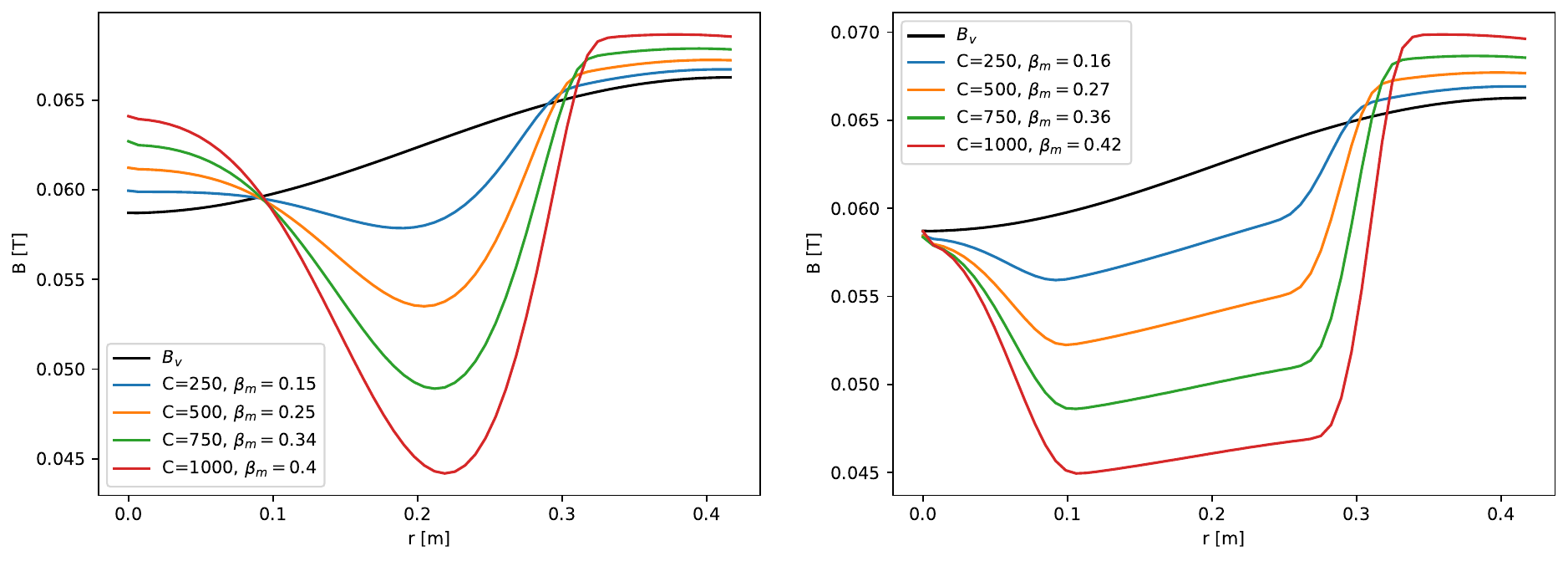}
    \caption{Diamagnetic field for a span of $C$ values for Eq. \eqref{eq:Novatron} with $A_n(\psi)=\psi_n(\psi - \psi_0)^2(\psi_1 - \psi)^2$ (\textit{left}) and $A_{flat}(\psi)$ (\textit{right}).}
    \label{fig:DiamagneticFields}
\end{figure}

An interesting aspect of magnetic mirror plasma equilibria in general is the induced magnetic field. It has been shown that the particle confinement time is increased when $\beta$ increases, and one of the reasons for this is the increase in mirror ratio \cite{beklemishev_diamagnetic_2016}. In Figure \ref{fig:DiamagneticFields} the diamagnetic field $B=B_v + B_p$ is shown for different equilibrium profiles. As can be seen, the plasma excavates the magnetic flux around itself, that is the plasma causes the total magnetic field near the axis and at the edges to increase. Furthermore, the plasma also causes the magnetic field to expand outwards. 

\section{Theoretical stability analysis}
There are three stability criteria in the MHD regime that should be taken into account: the mirror, fire-hose, and interchange instabilities. The first two are local instabilities, derived by Hastie and Taylor \cite{hastie_stability_1964, taylor_maximum_1965};
\begin{align}
    B/\mu_0 + d p_{\perp}/d B \geq 0, ~~~ B/\mu_0 - d p_{\parallel}/d B \geq 0,
\end{align}
where $B$ is the total magnetic field. These stability conditions are satisfied in all equilibrium profiles presented here. However, they do set a limit on the maximum $\beta_c$ of the equilibrium profile.

The interchange instability is, however, a global instability, related to the average curvature of the magnetic field along flux tubes with plasma. This instability arises in regions with ``bad'' curvature, that is the magnetic field line curvature points in the same direction as the pressure gradient.

In this section, the classic anisotropic Rosenbluth \& Longmire \cite{rosenbluth_stability_1957} criterion for interchange stability, based on particle orbit theory, is modified to be applicable to the Novatron's potentially slightly hollow pressure profile. Additionally, two novel fluid interchange stability criteria for \textit{anisotropic} plasmas are derived, based on anisotropic ideal MHD, and the double-adiabatic, Chew-Goldberger-Low (CGL) models, respectively.  A large exposition on the topic of the interchange criteria can be found in R. J. Hastie's MSc. thesis (1962) \cite{hastie_interchange_1962}.

\subsection{Modified Rosenbluth \& Longmire (MRL) criterion}

Rosenbluth \& Longmire \cite{rosenbluth_stability_1957} derived two interchange stability criteria at low $\beta$; one for isotropic plasmas, based on ideal MHD, and one for anisotropic plasmas, based on particle orbit theory. The latter criterion assumes that the magnetic moment $\mu$ and the longitudinal adiabatic invariant $J$ are both conserved. In Ref. \cite{jaderberg_introducing_2023} the validity of these assumptions for the Novatron is investigated numerically. The original Rosenbluth \& Longmire stability criterion, as found from particle orbit theory, reads 
\begin{align}
    \int{\frac{(p_{\parallel} + p_{\perp})}{RrB^2}}d\ell > 0, \label{eq:RnL}
\end{align}
where $r$ is the radial coordinate and $R$ is the field line curvature radius. In this formulation, $R$ is defined as positive ``if the center of curvature lies outside the plasma". If $R$ is positive everywhere along the field line, the integral will be positive. This definition assumes the pressure gradient always being directed into the plasma. 

Consider the curvature vector $\kappa$, which is directed towards the center of curvature, irrespective of the plasma's location. Thus, according to Rosenbluth \& Longmire, positive $R$ should be interpreted as the pressure gradient being directed anti-parallel to $\kappa$. Hence, using this definition instead, we can apply the criterion to cases where the direction of the pressure gradient points inwards. We can thus apply the Rosenbluth \& Longmire criterion also on the inner side of the plasma in the Novatron. The criterion, with $p_{tot}=(p_\parallel+p_\perp)/2$, then becomes:
\begin{align}
\int{-\text{sign(}\vec{\kappa} \cdot \nabla p_{tot}) \frac{p_{tot} |\vec{\kappa}|}{rB^2}} d \ell > 0. \label{eq:RnL_criterion}
\end{align}
We denote this criterion the Modified Rosenbluth \& Longmire (MRL) criterion.

\subsection{Generalized Rosenbluth \& Longmire Interchange (GRLI) criterion} \label{sec:GRLI}

We will now derive a novel interchange stability criterion, based on anisotropic ideal MHD. This generalized Rosenbluth \& Longmire criterion does not require the adiabatic moments $\mu$ and $J$ to be conserved, as is the case for the (particle orbit theory-based) MRL criterion. Furthermore the criterion takes, as opposed to isotropic ideal MHD, into account pressure variations along the magnetic field lines.

The ideal MHD model is derived in the collision-dominated limit where particle collision times are shorter than other typical time scales. Further, isotropization times for parallel and perpendicular ion pressures differ only by an order of magnitude from the ion-ion collision time $\tau_{ii}$ in the absence of anisotropic driving forces. Consequently, the plasma is considered isotropic in ideal MHD. However, to neglect resistivity, ideal MHD must assume $\tau_{ii}$ to be finite. Thus, anisotropic effects generated on short time scales, for example by strong NBI and RF heating, or by mirror confinement, could then consistently be included in the model without violating other assumptions \cite{shi_ideal_2016, brunetti_anisotropy_2020}.

To find a criterion for interchange stability we will now, along the lines of Rosenbluth and Longmire, investigate the circumstances under which it is energetically beneficial for the plasma to interchange flux tubes. For more details, please see the Appendix.

The (adiabatic) ideal MHD energy equation is
\begin{align}
    \frac{d}{dt}\Big(p \rho^{-\gamma}\Big) = 0, \label{eq:MHD}
\end{align}
where $\rho$ is the charge density. After integration we may write, where $'$ denotes a later time $t=t'$,
\begin{align}
\frac{p}{\rho^{\gamma}} = \frac{p'}{\rho'^{\gamma}}. \label{eq:MHD_time} 
\end{align}

Consider now two infinitesimally adjacent magnetic flux tubes, denoted by subscript 1 and 2, reaching all the way to the expander regions of the device. At time $t$ flux tube 1 is at position A and flux tube 2 is at position B. We assume that at time $t=t'$ the two flux tubes are interchanged, so that flux tube 1 is at position B while flux tube 2 is at position A. Assuming negligible induced fields ($\nabla \times \textbf B = \textbf 0$), the flux in the two flux tubes is the same, that is $\phi_1 = \phi_2$ in order to neglect the potentially stabilizing effect of field line bending, as shown by Rosenbluth \& Longmire \cite{rosenbluth_stability_1957}. As a consequence, the change in magnetic energy induced by the interchange is zero for both flux tubes. The perturbation in the radial direction however corresponds to a change in the magnetic flux function $\delta \psi$, where $\psi$ is associated with the total flux as a function of radius through the symmetry plane. 

We express local changes in terms of two infinitesimal cylindrical sections of the flux tubes with area $A$, height $d\ell$, volume $V$, and pressure $p$. During the interchange, these infinitesimal flux tubes are assumed to interchange volumes and magnetic fields.

The internal (or material) plasma energy density is
\begin{align}
    E_p = \frac {3}{2}(n_ek_BT_e + n_ik_BT_i) = 3nk_BT = \frac {3}{2} p,
\end{align}
for equal ion and electron densities and temperatures, with $p = (p_{\parallel} + p_{\perp})/2$. The material energy of an infinitesimal flux tube with height $d\ell$ can thus be written 
\begin{align}
    dE = \frac {3}{2} pV. 
\end{align}

To arrive at a criterion for interchange stability, we would like to express the total change in energy associated with the interchange of the two entire flux tubes. Since the change in magnetic energy induced by the interchange is zero, this total change in energy is equal to the total change in material energy, $\Delta E$.

We will begin by expressing the local changes where the change in energy for the two infinitesimal flux tubes $\Delta (dE) = \Delta (dE_1) +\Delta (dE_2)$ is then related to the stability criterion as follows:
\begin{align}
	\Delta E = \Delta \int dE = \int \Delta (dE) > 0,
\end{align}
and the integral sums up the changes in material energy of the infinitesimal flux cylinders along a flux tube.

We write, omitting the factor $3/2$,
\begin{align}
    & \Delta( dE_{1}) = p_1'V_1' - p_1V_1 = p_1'V_2 - p_1V_1, \label{eq:deltaE1} \\
    & \Delta (dE_{2}) = p_2'V_2' - p_2V_2 = p_2'V_1 - p_2V_2. \label{eq:deltaE2}
\end{align}
Let us now rewrite $p_1'$ using Eqs. (\ref{eq:MHD_time}):
\begin{align}
p_1' = p_1 \Big( \frac{\rho_1'}{\rho_1} \Big)^{\gamma} = p_1 \Big( \frac{V_1}{V_1'} \Big)^{\gamma} = p_1 \Big( \frac{V_1}{V_2} \Big)^{\gamma}, \label{eq:p_1_bis}
\end{align}
and similarly for $p_2'$ obtaining 
\begin{align}
p_2' = p_2 \Big( \frac{V_2}{V_1} \Big)^{\gamma}. \label{eq:p_2_bis}
\end{align}
Insertion of Eqs. \eqref{eq:p_1_bis} and \eqref{eq:p_2_bis} into Eqs. \eqref{eq:deltaE1} and \eqref{eq:deltaE2} yields
\begin{align}
& \Delta( dE_{1}) = p_1 \Big( \frac{V_1}{V_2} \Big)^{\gamma} V_2 - p_1V_1, \label{eq:deltaE1_2} \\
& \Delta( dE_{2}) =  p_2 \Big( \frac{V_2}{V_1} \Big)^{\gamma} V_1 - p_2V_2. \label{eq:deltaE2_2}
\end{align}

Next, we rewrite the pressure, volume and magnetic field strength for flux tube 2 in terms of those for flux tube 1 and obtain
\begin{align}
    p_2 = p_1 + \delta p, ~~~ V_2 = V_1 + \delta V. \label{eq:p_v_delta}
\end{align}

Inserting this into Eqs. \eqref{eq:deltaE1_2} and \eqref{eq:deltaE2_2}, keeping terms up to 2nd order and setting $p_1=p$ and $V_1=V$, we can write (see Appendix for details) 
\begin{align}
    \Delta (dE) &= \Delta (dE_1) +\Delta (dE_2) \\
    &= (\gamma-1) \Bigg(\delta p + \gamma p \frac{\delta V}{V}\Bigg)\delta V  \label{eq:deltaEp}
\end{align} 

As mentioned above, to arrive at a criterion, we need to add the contributions from all infinitesimal flux tubes. We note that the flux $\phi = AB$ is constant along the flux tube so that
\begin{align}
     V = A d\ell = \phi \frac{d\ell}{B}. \label{eq:phi}
\end{align}

Using that for stability it is required that
\begin{align}
	\Delta E = \Delta \int dE = \int \Delta (dE) > 0  ,
\end{align}
we obtain (for details see Appendix): 
\begin{align}
	\int \delta \Bigg(\frac {1}{B}\Bigg) B^{2 \gamma} \delta \Bigg( \frac {p_{\parallel} + p_{\perp}} {2B^{2 \gamma}}\Bigg) d\ell > 0  .
\end{align}

This is the Generalized Rosenbluth \& Longmire Interchange (GRLI) criterion. Here the functions, on which $\delta$ operate, are non-constant along the field lines. For practical application, we may assume the paraxial limit in which $|d/dr| \gg |d/dz|$ and introduce the flux coordinate $\psi$, arriving at the approximate stability condition
\begin{align}
    \int \frac{\partial}{\partial \psi}\Bigg(\frac{1}{B}\Bigg)B^{2\gamma}\frac{\partial}{\partial \psi}\Bigg(\frac{p_{\parallel} + p_{\perp}}{2B^{2\gamma}}\Bigg) d\ell > 0. \label{eq:GRLI_criterion}
 \end{align}

Clearly, a sufficient condition for stability is obtained if the integrand is everywhere positive, that is
\begin{equation}
	\frac {\partial}{\partial \psi} \Bigg(\frac {1} {B}\Bigg) \ \frac {\partial}{\partial \psi} \Bigg(\frac {p_\parallel + p_\perp} {2B^{2 \gamma}}\Bigg) > 0.
\end{equation}

For stability in a magnetic well, with $(\partial/\partial \psi)(1/B)< 0$, $p$ should thus decrease with $\psi$ or increase slower than $B^{2 \gamma}$. The marginal, sufficient condition becomes
\begin{equation}
	\frac {\partial}{\partial \psi} \Bigg(\frac {p_\parallel + p_\perp} {2B^{2 \gamma}}\Bigg) = 0 ,
\end{equation}
or 
\begin{equation}
	p_\parallel + p_\perp = C  {B^{2 \gamma}},
\end{equation}
where $C$ is a constant.

One should note that the criterion Eq. \eqref{eq:GRLI_criterion} allows for a stronger variation of $p$ with respect to $\psi$, and thus a more hollow pressure profile.


\subsection{Chew-Goldberger-Low Interchange (CGLI) criterion} \label{sec:CGLI}

The interchange stability criterion (\ref{eq:GRLI_criterion}) is general, however the criterion is based on collisional MHD and fusion plasmas are typically collisionless. We will now, for comparison, derive a corresponding interchange stability criterion based on the collisionless Chew-Goldberger-Low (CGL) fluid model \cite{chew_boltzmann_1956, hunana_introductory_2019, kaur_magnetothermodynamics_2019, le_hybrid_2016}. Some details of the derivation of this Chew-Goldberger-Low Interchange (CGLI) criterion can be found in the Appendix. 
 
The CGL model assumes vanishing heat flux, and features the following adiabatic, anisotropic equations of state:
\begin{align}
    &\frac{d}{dt}\Bigg(\frac{p_{\parallel}B^2}{\rho^3}\Bigg) = 0, \\
    &\frac{d}{dt}\Bigg(\frac{p_{\perp}}{\rho B}\Bigg) = 0, 
\end{align}
or, using similar terminology as in Sec.~\ref{sec:GRLI}, 
\begin{align}
\frac{p_{\parallel}B^2}{\rho^3}=\frac{p_{\parallel}'B'^2}{\rho'^3}, \label{eq:dad_1} \\
\frac{p_{\perp}}{\rho B}=\frac{p_{\perp}'}{\rho' B'},  \label{eq:dad_2}
\end{align}
valid for any infinitesimal plasma fluid element. Consider two adjacent flux tubes, as defined in Sec. \ref{sec:GRLI}, but where the pressures are now governed by (\ref{eq:dad_1}) and (\ref{eq:dad_2}). Using Eqs. \eqref{eq:dad_1} and \eqref{eq:dad_2}, valid for both flux tube 1 and 2 separately, we obtain
\begin{align}
    p_1' &= \frac{1}{2}(p_{\parallel, 1}' + p_{\perp,1}') \nonumber \\
    &=\frac{1}{2}\Bigg(\frac{p_{\parallel,1}B_1^2V_1^3}{B_1'^2V_1'^3} + \frac{p_{\perp 1}V_1B_1'}{V_1'B_1}\Bigg), \label{eq:p_bis_1}
\end{align}
and a similar expression for $p_2'$. We now (see Appendix for details) substitute  Eq.\eqref{eq:p_bis_1} (and $p_2'$ equivalent) into Eqs. \eqref{eq:deltaE1} and \eqref{eq:deltaE2} to obtain
\begin{align}
  \Delta (dE) &= \Delta (dE_1) +\Delta (dE_2) \nonumber \\
    &=\frac{1}{2}\Bigg[\Bigg(\frac{p_{\parallel, 1}B_1^2V_1^3}{B_2^2V_2^3} + 
     \frac{p_{\perp, 1}V_1B_2}{V_2B_1}\Bigg)V_2 \nonumber \\
     &-(p_{\parallel 1} + p_{\perp 1})V_1 \nonumber \\
    &+\Bigg(\frac{p_{\parallel 2}B_2^2V_2^3}{B_1^2V_1^3} + 
    \frac{p_{\perp 2}V_2B_1}{V_1B_2}\Bigg)V_1 \nonumber \\
    &- (p_{\parallel 2} + p_{\perp 2})V_2\Bigg]. \label{eq:delta_E_prim_only}
\end{align}

Next, the pressures, volumes, and magnetic field strength for flux tube 2 are rewritten in terms of those for flux tube 1:
\begin{align}
    p_{\parallel 2} = p_{\parallel 1} + \delta p_{\parallel}, \label{eq:p_papa_delta} \\
    p_{\perp 2} = p_{\perp 1} +  \delta p_{\perp}, \label{eq:p_perp_delta} \\
    V_{2} = V_{1} + \delta V, \label{eq:v_delta} \\
    B_{2} = B_{1} + \delta B. \label{eq:B_delta}
\end{align}
Substituting these equations into Eq. \eqref{eq:delta_E_prim_only}, setting $p=p_1$, and keeping terms up to 2nd order we obtain after some algebra (see Appendix for details):
\begin{align}
    \Delta E \propto \bigintsss\delta \Bigg(\frac{1}{B}\Bigg) \Bigg[B^4\delta\Bigg(\frac{p_{\parallel}}{B^4}\Bigg) + \frac{B^3}{2}\delta\Bigg(\frac{p_{\perp}}{B^3}\Bigg)\Bigg]d\ell > 0. \label{eq:CGL_criterion}
\end{align} 

We denote this criterion as the Chew-Goldberger-Low Interchange (CGLI) criterion. Again, using the flux coordinate $\psi$, the stability condition is approximated by
\begin{align}
    \bigintsss \frac{\partial}{\partial \psi} \Bigg(\frac{1}{B}\Bigg) \Bigg[B^4\frac{\partial}{\partial \psi}\Bigg(\frac{p_{\parallel}}{B^4}\Bigg) + \frac{B^3}{2}\frac{\partial}{\partial \psi}\Bigg(\frac{p_{\perp}}{B^3}\Bigg)\Bigg]d\ell > 0. \label{eq:CGLI_criterion}
\end{align}

\subsection{Stability results}
\begin{figure}[H]
    \centering
    \includegraphics[width=1.0\textwidth]{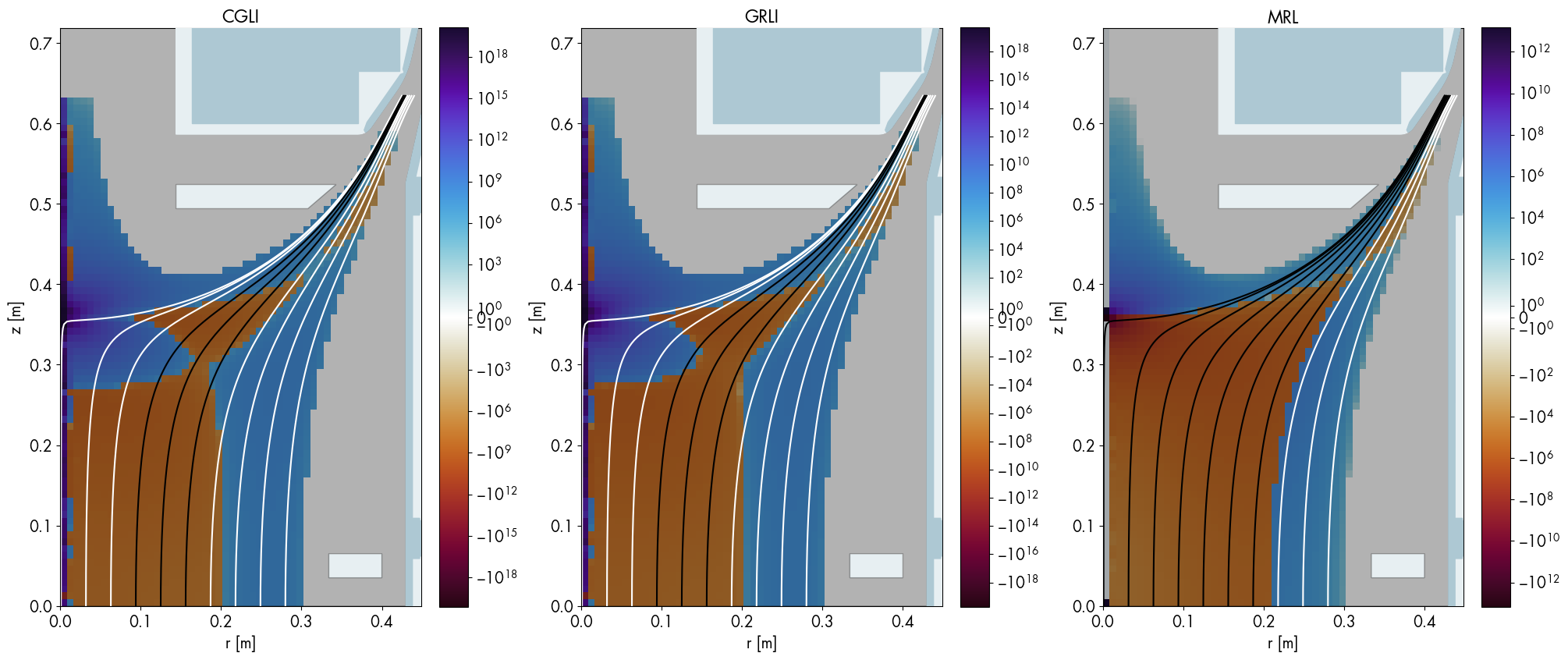}
    \caption{Symlog plots presenting stability regions in the plasma according to the three interchange criteria: CGLI, GRLI, and MRL. Stable flux lines are shown as white and unstable as black. The anisotropic pressure profiles are Eqs. \eqref{eq:Novatron} and \eqref{eq:norm_fisher_killen}, with parameters $B_1 = 0.34$, $\psi_0 = -0.00023$, $\psi_1 = 0.0028$, $M=3$, $N=2$, $C = 100$, and $\langle{\beta}\rangle_V=0.033$.}
    \label{fig:InterchangeCriteriaKillenNovatron}
\end{figure}

\begin{figure}[H]
    \centering
    \includegraphics[width=1.0\textwidth]{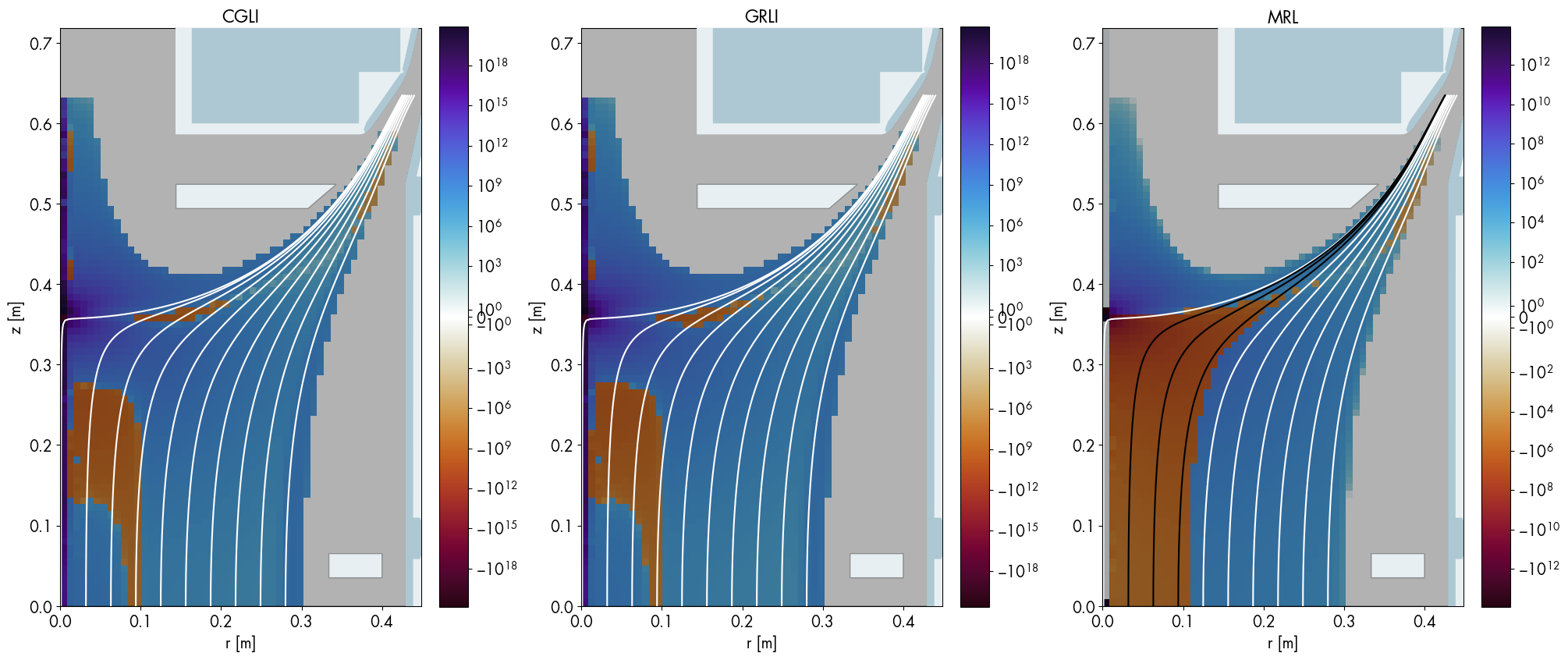}
    \caption{As Fig. \ref{fig:InterchangeCriteriaKillenNovatron}, here with the anisotropic pressure profiles Eqs. \eqref{eq:Novatron} and $A_{flat}(\psi)$, with parameters $B_1 = 0.34$, $\psi_0 = -0.00023$, $\psi_1 = 0.0028$, $M=3$, $C = 100$, and $\langle{\beta}\rangle_V=0.057$.}
    \label{fig:InterchangeCriteriaFlatNovatron}
\end{figure}

The anisotropic equilibrium pressure profiles obtained from the mirror equilibrium equations have been applied to the three interchange criteria above. The results can be seen in Figures \ref{fig:InterchangeCriteriaKillenNovatron} and \ref{fig:InterchangeCriteriaFlatNovatron}, which show low volume-average $\beta\equiv\langle\beta\rangle_V$ (see definition of the local $\beta$ in Eq. \eqref{eq:local_beta}) pressure profiles; one that features regions of unstable flux tubes, and another where all flux tubes throughout the plasma are stable, respectively. Both figures show regions where positive values identify the regions of stability, while negative values identify regions of instability. Since the interchange criteria are field line integrals over the shown criteria integrands, a set of field lines have also been plotted to show which magnetic flux tubes are stable. In the Figures \ref{fig:InterchangeCriteriaKillenNovatron} and \ref{fig:InterchangeCriteriaFlatNovatron} the stable flux lines are shown as white and unstable as black.

Figure \ref{fig:InterchangeCriteriaKillenNovatron} is an example of an equilibrium pressure profile featuring stable flux tubes at the edge of the plasma, followed by a set of unstable flux tubes as the plasma gradient decreases towards the axes. A key feature of the two additional interchange criteria Eqs. \eqref{eq:GRLI_criterion} and \eqref{eq:CGLI_criterion}, when applied to the Novatron, is that the flux tubes closer to the axes are stable. Figure \ref{fig:InterchangeCriteriaFlatNovatron} shows an example where all flux tubes are stable, as obtained from the CGLI and GRLI criteria, throughout the entire plasma region. Thus the anisotropic equilibrium pressure profiles are MHD stable.

A noteworthy aspect of the Novatron stability analysis is the convex curvature present on the outer flux lines. This curvature led pioneering plasma theorists, such as Harold Grad and his contemporaries, to assert the mirror-cusp’s MHD stability \cite{grad_theory_1957, bishop_project_1958, taylor_stable_1963, haines_plasma_1977}. The reasoning was that the plasma is encased within a zone of favorable curvature. This argument is consistent with the Novatron’s MHD stability properties. However, the inner plasma region, proximal to the axes, may undergo flux line interchange if the plasma gradients become excessively steep. The outcome of this type of interchange is a mixing and leveling of the density profiles. Importantly, as seen from PIC simulations,this process does not lead to global plasma instability, but rather serves to redistribute the plasma within the system.

\section{WarpX: Particle-in-cell simulations}
To verify the global stability properties of the Novatron, the GPU-accelerated, massively parallel Particle-In-Cell (PIC) code WarpX has been utilized \cite{fedeli_pushing_2022}. WarpX is an open-source initiative primarily supported by the US DOE Exascale Computing Project and builds on the AMReX block-structured mesh-refinement framework \cite{zhang_amrex_2019}. WarpX is designed as a fully and hybrid kinetic electromagnetic PIC code, accommodating a wide range of user inputs, such as initial and external fields, multiple species, embedded boundaries, and various physics modules, among others. To enhance the WarpX framework, we have integrated support for multiple-cut polygon geometries, imported from a standard text format (WKT). This enables modeling of a Novatron vacuum vessel and limiters as embedded boundaries.

\subsection{Fully kinetic PIC}
The governing equations of the Particle-in-Cell method for plasma simulations are based on the collisionless Vlasov equation,
\begin{align}
    \frac{\partial f_s}{\partial t} + \mathbf{v} \cdot \nabla_x f_s + \frac{q_s}{m_s}(\mathbf{E} + \mathbf{v} \times \mathbf{B}) \cdot \nabla_v f_s = 0,
    \label{eq:vlasov}
\end{align}
where $f_s(\mathbf{x}, \mathbf{v}, t)$ is a 6-dimensional distribution function, specific for each particle species $s$, $x$ is the spatial coordinate, $\mathbf{v}$ is the velocity space coordinate, $q_s$ is the charge, and $m_s$ is the mass. The electric and magnetic fields, $\mathbf{E}$ and $\mathbf{B}$, are described by Maxwell's equations.

One way to solve Eq. \eqref{eq:vlasov} is to discretize all dimensions on a computational grid where the derivatives can be computed with a numerical method of choice (for example finite difference). While this method is straightforward, it becomes computationally impractical without further approximations or dimensional reductions.

The Particle-in-Cell (PIC) method addresses this challenge by introducing macro-particles. These macro-particles enable the reconstruction of the distribution function, with each individual particle's motion governed by the Newton-Lorentz force equations,
\begin{align}
    \frac{d\mathbf{x}_i}{dt} = \mathbf{v}_i, ~~~ \frac{d\mathbf{v}_i}{dt} = \frac{q_i}{m_i}(\mathbf{E}(x_i) + \mathbf{v}_i \times \mathbf{B}(x_i)),
    \label{eq:newton-lorentz}
\end{align}
where $\mathbf{v}_i$ is the velocity of the $i$th particle, $q_i$ is the charge, and $m_i$ is the mass of the particle. The particle dynamics is coupled to Faraday's law of induction, 
\begin{align}
    \nabla \times \mathbf{E} = -\frac{\partial \mathbf{B}}{\partial t},
\end{align}
and Ampere's Law with Maxwell's addition,
\begin{align}
    \nabla \times \mathbf{B} = \mu_0\mathbf{J} + \mu_0\varepsilon_0\frac{\partial \mathbf{E}}{\partial t},
\end{align}
where $\mathbf{J}$ is the current density, $\varepsilon_0$ is the vacuum permittivity, and $\mu_0$ is the vacuum permeability. Thus, the particles create self-consistent electromagnetic fields as they move through space. 

The main drawback of the fully kinetic PIC, when applied to multi-scale plasma physics simulations, is that the grid resolution needs to be smaller, or close to, the Debye length. The issue is then compounded when using explicit time-stepping methods that need to use excessively small time-steps to converge and satisfy the CFL criteria.

\subsection{Hybrid-PIC}
Transitioning from a fully kinetic PIC approach, where both electrons and ions are treated kinetically, to a \emph{hybrid-PIC} means treating the electrons as a fluid while the ions are still treated kinetically \cite{le_hybrid-vpic_2023, groenewald_accelerated_2023}. The hybrid-PIC method has a long history in plasma physics simulations, and has been used for problem-solving in many physical systems, for example magnetic mirrors \cite{kumar_kinetic_2022, le_hybrid-vpic_2023}, magnetic reconnection \cite{karimabadi_magnetic_2004, le_hybrid_2016, finelli_bridging_2021}, Hall-thrusters \cite{vchivkov_optimization_2004, villafana_3d_2023, petronio_study_2024}, Spherical Tokamaks \cite{lestz_hybrid_2021, belova_coupling_2015}, and astrophysics \cite{kunz_firehose_2014}. It has been shown, by comparing to a fully kinetic PIC model, that the hybrid-PIC method can accurately solve plasma dynamics when a proper equation of state for the electrons are tailored to the specific problem being solved \cite{le_hybrid_2016, finelli_bridging_2021}.

Here we provide a brief outline for how the hybrid-PIC method functions. Since the ions are still treated kinetically, their motion is governed by the Newton-Lorentz force Eqs. \eqref{eq:newton-lorentz}, and thus their function in the model remains unchanged compared to their function in the fully kinetic version. The electron dynamics, however, now described as a fluid, is governed by the conservation of momentum,
\begin{align}
    m_en_e\frac{d \mathbf{u}_e}{d t} = -\nabla \cdot \mathbf{P}_e - en_e(\mathbf{E} + \mathbf{u}_e \times \mathbf{B}) + \mathbf{R}_{ei},
\end{align}
where $\mathbf{u}_e$ is the electron fluid velocity, $\mathbf{P}_e$ is the electron pressure tensor, and $\mathbf{R}_{ei}$ is an electron-ion collision term. 

The trick in the hybrid-PIC method is that the electron fluid velocity is not directly computed. Instead the electric field is isolated as follows,
\begin{align}
    \mathbf{E} = -\frac{1}{en_e}\nabla \cdot \mathbf{P}_e - \frac{m_e}{e}\frac{d \mathbf{u}_e}{d t} -  \mathbf{u}_e \times \mathbf{B} + \frac{1}{en_e}\mathbf{R}_{ei}.
\end{align}
The electron dynamics are then linked to the ions through the total current density, where quasi-neutrality ($n=n_i Z = n_e$, where $Z$ is the charge number), has been assumed,
\begin{align}
    \mathbf{J} = en(\mathbf{u}_i - \mathbf{u}_e) = \mathbf{J}_i - en \mathbf{u}_e.
\end{align}
In this description, the plasma is also assumed to be non-relativistic, i.e. $\epsilon_0 \rightarrow 0$ and the Maxwell-Ampère equation becomes $\mu_0 \mathbf{J} = \nabla \times \mathbf{B}$. This gives us,
\begin{align}
    \mathbf{u}_e = \frac{1}{en}(\mathbf{J}_i - \mathbf{J}) = \frac{1}{en}\Bigg(\mathbf{J}_i - \frac{1}{\mu_0}\nabla \times \mathbf{B}\Bigg).
\end{align}
Another common approximation used in the hybrid-PIC method is to set the electrons as a mass-less fluid, that is $m_e=0$, which simplifies their description to a generalized Ohm's law \cite{le_hybrid-vpic_2023, groenewald_accelerated_2023},
\begin{align}
    \mathbf{E} = \frac{1}{en}\Bigg[\Bigg(\frac{1}{\mu_0}\nabla \times \mathbf{B} - \mathbf{J}_i\Bigg)\times\mathbf{B} -\nabla p_e\Bigg] + \eta \mathbf{J} - \eta_H \nabla^2 \mathbf{J}\label{eq:Ohm}
\end{align}    
where $p_e$ is the electron pressure, here treated as a scalar, $n$ is the number density, and $e$ is the elementary charge. Here the collision term $\mathbf{R}_{ei}/en = \eta \mathbf{J} - \eta_H \nabla^2 \mathbf{J}$ has been used, where $\eta$ is the resistivity and $\eta_H$ is the hyper-resistivity.

The last unknown variable needed in order to close the system is the electron pressure. As with most fluid descriptions of plasma, the equation of state (EOS) is an ad-hoc addition. As of writing, WarpX supports the isotropic EOS,
\begin{align}
    p_e = p_0\Bigg(\frac{n_e}{n_0}\Bigg)^{\gamma},
    \label{eq:electron_pressure}
\end{align}
where $\gamma=1$ is the iso-thermal case, $p_0 = n_0 T_{e0}$ is the reference electron pressure, and $\gamma=5/3$ is the adiabatic case.

To summarize, by treating electrons as a fluid, the hybrid-PIC approach reduces computational complexity while still capturing essential ion kinetics. This method is particularly useful in scenarios where ion dynamics are of primary interest, and electron thermal effects can be modeled without resolving their individual particle trajectories, for instance when studying macro instabilities \cite{groenewald_accelerated_2023}.

\subsection{Simulations}
For this work, we have extended the WarpX hybrid-PIC code to allow for the inclusion of a static external magnetic field.  This is done by splitting the total magnetic field as $\mathbf{B} = \mathbf{B}_v + \mathbf{B}_p$, where $\mathbf{B}_v$ is the static external magnetic field and $\mathbf{B}_p$ is the time-varying plasma induced magnetic field. A few terms can be neglected in the process, since $\nabla \times \mathbf{B}_v = \mathbf{0}$ and $\partial \mathbf{B}_v /\partial t = \mathbf{0}$. The result is that the static external field only enters the generalized Ohm's law Eq. \eqref{eq:Ohm} and in the evolution of the ion velocities, Eq. \eqref{eq:newton-lorentz}.

To begin with, we are interested in demonstrating the difference in interchange stability between a classical magnetic mirror and a first-generation Novatron. The classical magnetic mirror is known to be unstable to interchange modes, assuming no stabilization techniques are employed, such as Finite-Larmor-Radius (FLR) or Vortex stabilization. Therefore, the hybrid-PIC code must be capable of capturing these modes and their growth rates within a reasonable margin of error. We can then proceed to investigate the plasma in a Novatron configuration. If the simulations show that these modes either do not occur or are dampened, we can conclude the stability of the Novatron configuration in comparison to a classic mirror (without stabilization techniques). In absence of any other significant differences between the two simulation runs, we will attribute any major differences to the variations in magnetic topology. 

There are several numerical considerations to be taken into account when running a hybrid-PIC code. The Hall term in the generalized Ohm's law gives rise to the strongest limit on the time step due to Whistler waves. The Whistler waves set a Courant-Friedrichs-Lewy (CFL) equivalent criterion, which sets a limit on the relation between spatial and temporal resolutions for a stable solution. The CFL criterion for Whistler waves is $\Omega_{ci}\Delta t < (\Delta x /d_i)^2$ \cite{le_hybrid-vpic_2023}, where $d_i = c / \Omega_{ci}$ is the ion skin depth and $\Omega_{ci} = ZeB / m_i$ is the ion cyclotron frequency. 

Another factor is the density floor selected for the simulation. To set the density floor, we use the reference density and adjust the density floor $n_f$ to a few percentages of it, that is, $n_f/n_0 \sim 1 - 5\%$, where $n_0$ is a user-defined reference density, used also in calculating the electron pressure Eq. \ref{eq:electron_pressure}. Here, we choose to set the reference density $n_0$ to the maximum of the initial ion density.

\subsubsection{Results}

A comparison between a classical mirror and a Novatron has been made. Since the configurations differ inherently in several ways it is not apparent which aspects of the geometry and temperature to equate, since some choices may speak in favor of one or the other concept in terms of stability. First, a classical mirror is more stable towards interchanges if it is long and thin, making the curvature less unfavorable. Second, a higher temperature will increase stability for the classical mirror as it enhances finite Larmor radius (FLR) effects. These features bear less difference for the Novatron, which by its favorable curvature should be inherently stable towards interchange modes. 

In this comparison the classical mirror's initial conditions have been chosen so as to include FLR stabilization, which has been shown to stabilize interchange modes with $m\geq2$. This can be done by satisfying the criteria derived  by Ryutov et. al (2011) \cite{ryutov_magneto-hydrodynamically_2011},
\begin{align}
    m > 2 \frac{a / \rho_i}{L_p / a},
\end{align}
where $a$ is the plasma radius, $\rho_i$ is the ion Larmor radius, and $L_p$ is the mirror half-length. In this simulation the parameters are $a / \rho_i = 6.7$ and $L_p / a = 7.1$, which satisfies the criteria.

A difference between the two configurations is the larger simulation domain chosen for the Novatron as extra space is required for the outer expanders, which are absent in the classical mirror case. The heights (along central axis) are, however, equal between the two simulations. The spatial resolution is then slightly lower for the Novatron since a larger simulation domain is required.

\begin{table}[h!]
\centering
\resizebox{0.7\textwidth}{!}{%
\begin{tabular}{ |l|c|c|  }
\hline
Simulation parameters & Classic mirror & Novatron \\
\hline
Simulation domain [m] & $x, y \in \pm 0.25,~z \in \pm1$ & $x, y \in \pm 0.7,~z \in \pm1$ \\
Number of cells & $92\times92\times368$ & $220\times220\times310$ \\
Time step [s] & $10^{-9}$ s & $10^{-9}$ s \\
Particle shape function & 2 & 2 \\
Initial $p$ profiles & Flat/Cutler & Flat/Novatron \\
$T_e$ [eV] & 20 & 20 \\
$T_i$ [eV] & 100 & 100 \\
Embedded boundary & on & on \\
$\eta$ [$\Omega \cdot \text{m}$] & $10^{-7}$ & $10^{-7}$ \\
$\eta_H$ [$\Omega \cdot \text{m}^3$] & $10^{-6}$ & $10^{-6}$ \\
$n_f/n_0$ & 5 \% & 5 \% \\
Filter & 1-pass & 1-pass \\
\hline
\end{tabular}
}
\caption{Parameters used for the simulations in the Classic mirror and Novatron configurations.}
\label{tab:warpx_simulation}
\end{table}

\begin{figure}[H]
    \centering
    \includegraphics[width=0.52\textwidth]{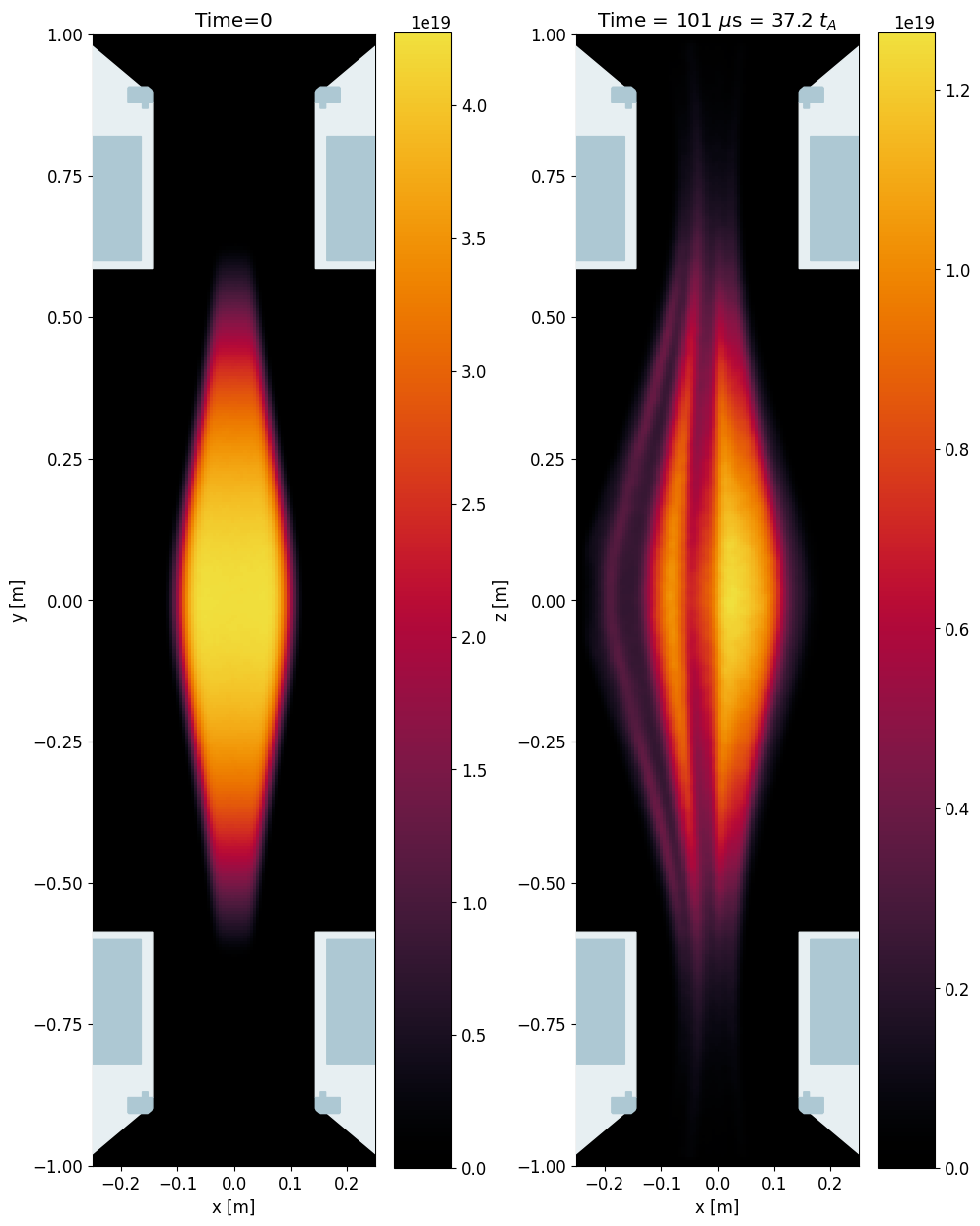}
    \caption{The initial and final ($t=101\ \mu \text{s} = 37.2\ t_A$) number density in the classical mirror.}
    \label{fig:CMinitfinaldensity}
\end{figure}

\begin{figure}[H]
    \centering
    \includegraphics[width=0.95\textwidth]{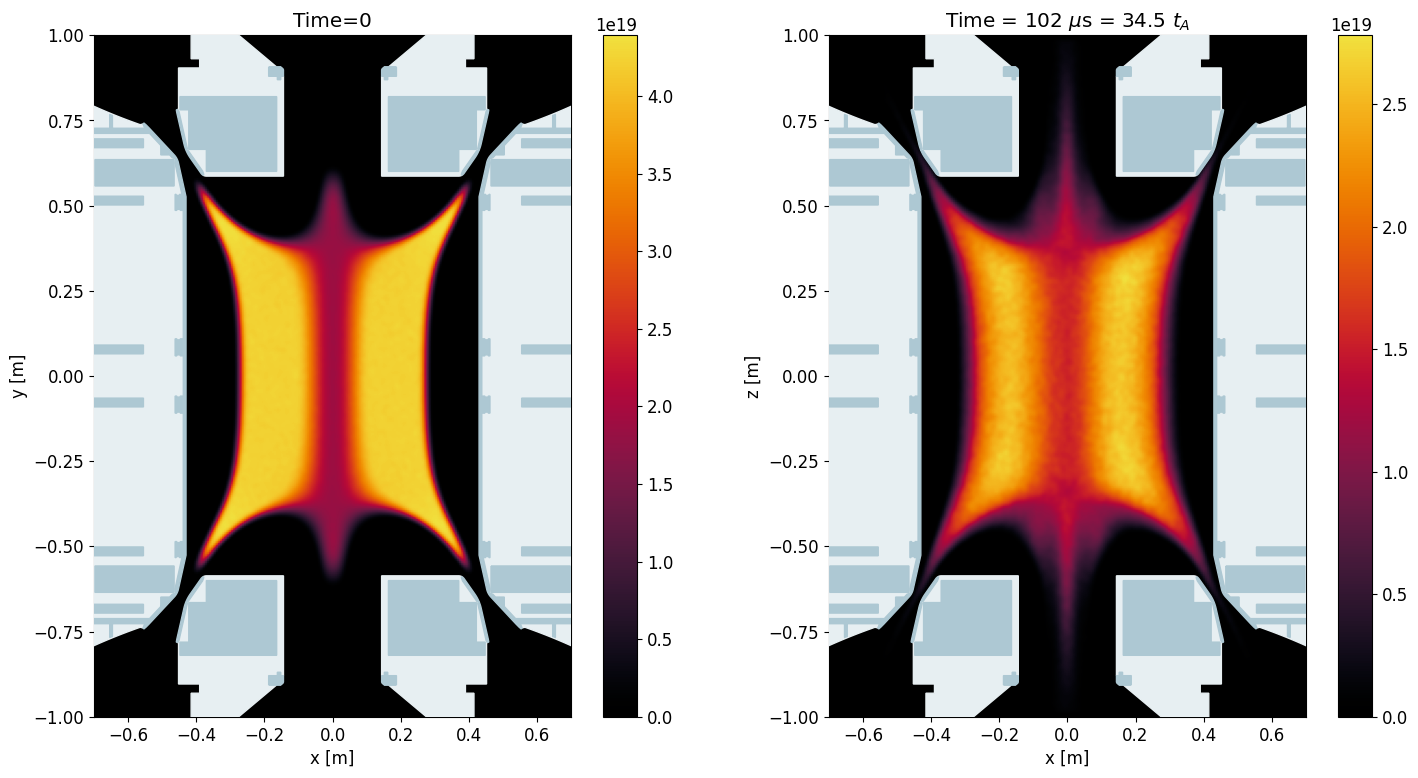}
    \caption{The initial and final ($t=102\ \mu \text{s} = 34.5\ t_A$) number density in the Novatron.}
    \label{fig:N1initfinaldensity}
\end{figure}

The initial conditions for the ions were derived from the Cutler Eq. \eqref{eq:cutler} pressure profiles for the classical mirror and the new pressure profile Eq. \eqref{eq:Novatron} for the Novatron. The initial and final number density for the classical mirror and Novatron can be seen in Figures \ref{fig:CMinitfinaldensity} and \ref{fig:N1initfinaldensity}, respectively. The simulation parameters used in both the classical mirror and Novatron simulation are presented in Table \ref{tab:warpx_simulation}. It should be noted that this simulation study is not meant to be a thorough parameter sweep and hence there could be a range of interesting parameters that are not covered here.

\begin{figure}[H]
    \centering
    \includegraphics[width=0.9\textwidth]{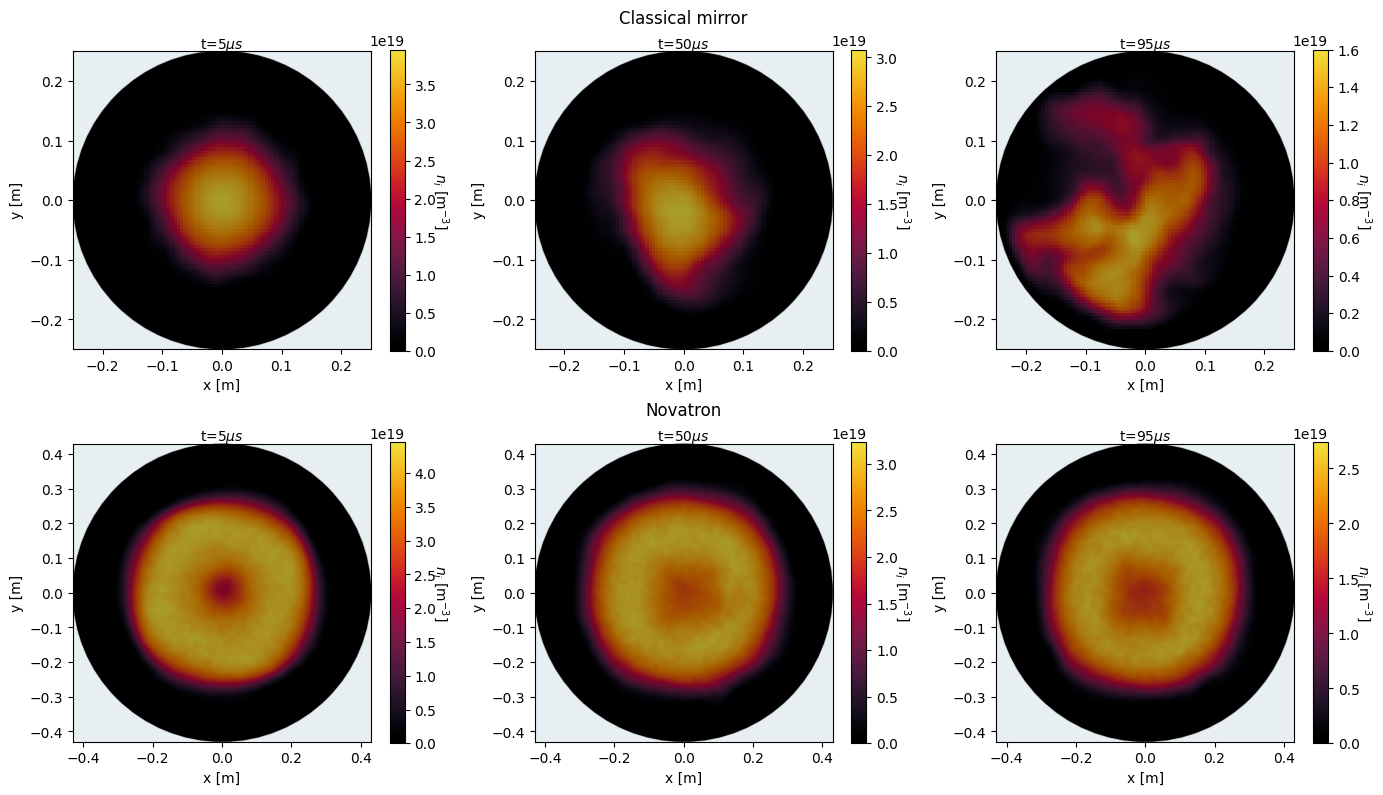}
    \caption{WarpX Hybrid-PIC simulation of a classical magnetic mirror (upper), and a Novatron (lower). The figures show the number density at times $t=5,\ 50,\ $ and $95\ \mu s$, left to right. This amounts to Alfvén times roughly $t = 1.84,\ 18.4,\ $ and $35\ t_A$ for the classical mirror and $t = 1.69,\ 16.9,\ $ and $32.1\ t_A$ for the Novatron.}
    \label{fig:RhoHybridWarpXNovatronClassicalTest2}
\end{figure}

\begin{figure}[H]
    \centering
    \includegraphics[width=0.9\textwidth]{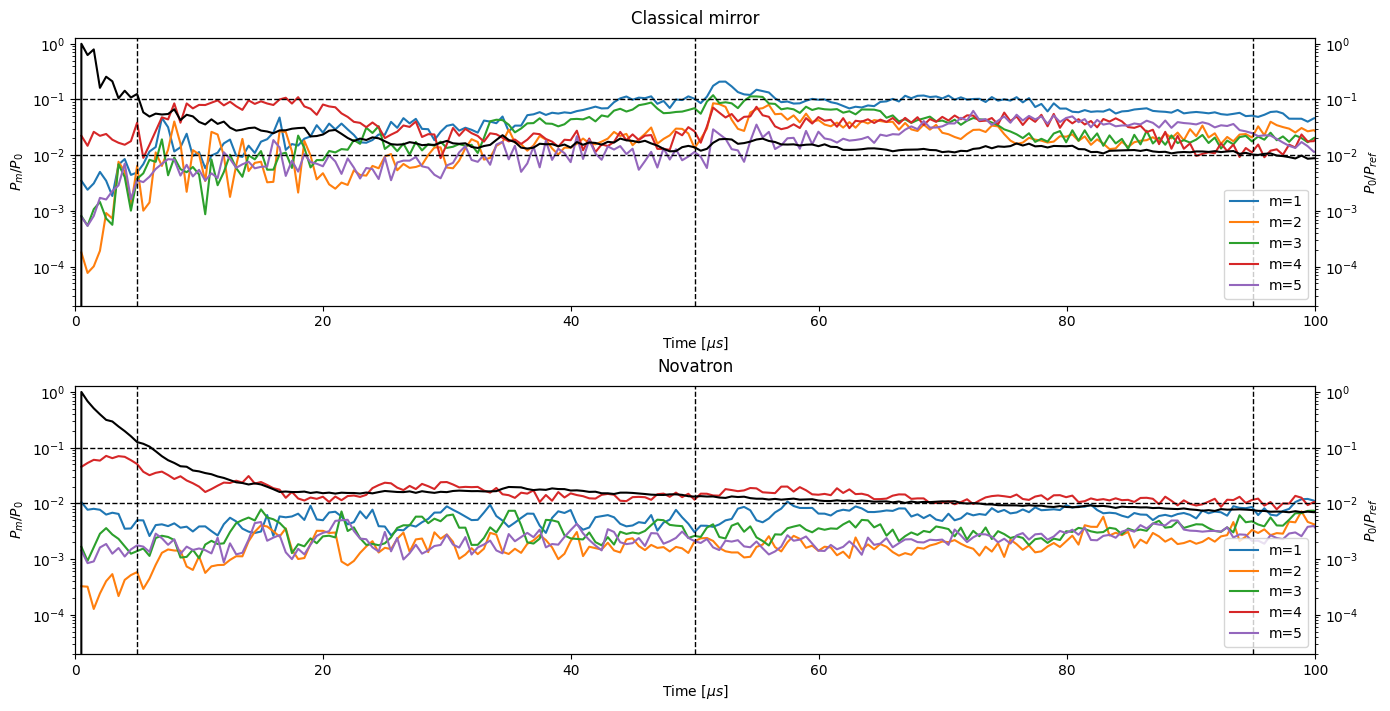}
    \caption{The time evolution of the power ratio $P_m/P_0$ of azimuthal modes ($m\leq5$) (resulting from an FFT analysis) of the electric field for the classical mirror (upper) and Novatron (lower).  The values are $r$ weighted radial averages taken at $z=0$, with dashed vertical lines indicating the timestamps of the number density seen in Figure \ref{fig:RhoHybridWarpXNovatronClassicalTest2}. The black lines (right axes) show the time-series of the normalized zeroth azimuthal mode power $\hat{P} = P_0 / P_{ref}$, where $P_{ref} = P_0(t=5\mu s)$.}
    \label{fig:AzimuthalsHybridWarpXNovatronClassicalTest2}
\end{figure}

The time-series of the number density results can also be seen at the mid-plane ($z=0$) for both devices in Figure \ref{fig:RhoHybridWarpXNovatronClassicalTest2}. Here, a dominant global $m=1$ mode has been established in the classical mirror. This result corroborates the theory that higher azimuthal modes are damped in a long, thin, and hot mirror plasma, and it gives further credence to the hybrid-PIC simulations. In Figure \ref{fig:AzimuthalsHybridWarpXNovatronClassicalTest2} the power ratio $P_m/P_0$ for the azimuthal modes $m\leq 5$ can be seen for the classical mirror and Novatron. The radial mean growth rate of the $m=1$ azimuthal mode was calculated as $\langle \gamma \rangle_r=0.04\ \mu s^{-1}$, giving a lower estimate of the growth rate. From the $m=1$ azimuthal mode at $r = a / 2 = 0.05$ m the growth rate was estimated as $\gamma=0.14\ \mu s^{-1}$. The theoretical value, taken from Ryutov et. al \cite{ryutov_magneto-hydrodynamically_2011}, is $\Gamma_0 \sim v_{Ti} / L =  0.12\ \mu s^{-1}$, where $L$ denotes the scale length for axial changes in the magnetic field. Here, $L$ is approximated as the mean value of $B / (dB/dz)|_{r=0}$, giving a value of $L=0.73$ (similar to the mirror half-length $L_p=0.71$).

\begin{figure}[H]
    \centering
    \includegraphics[width=0.9\textwidth]{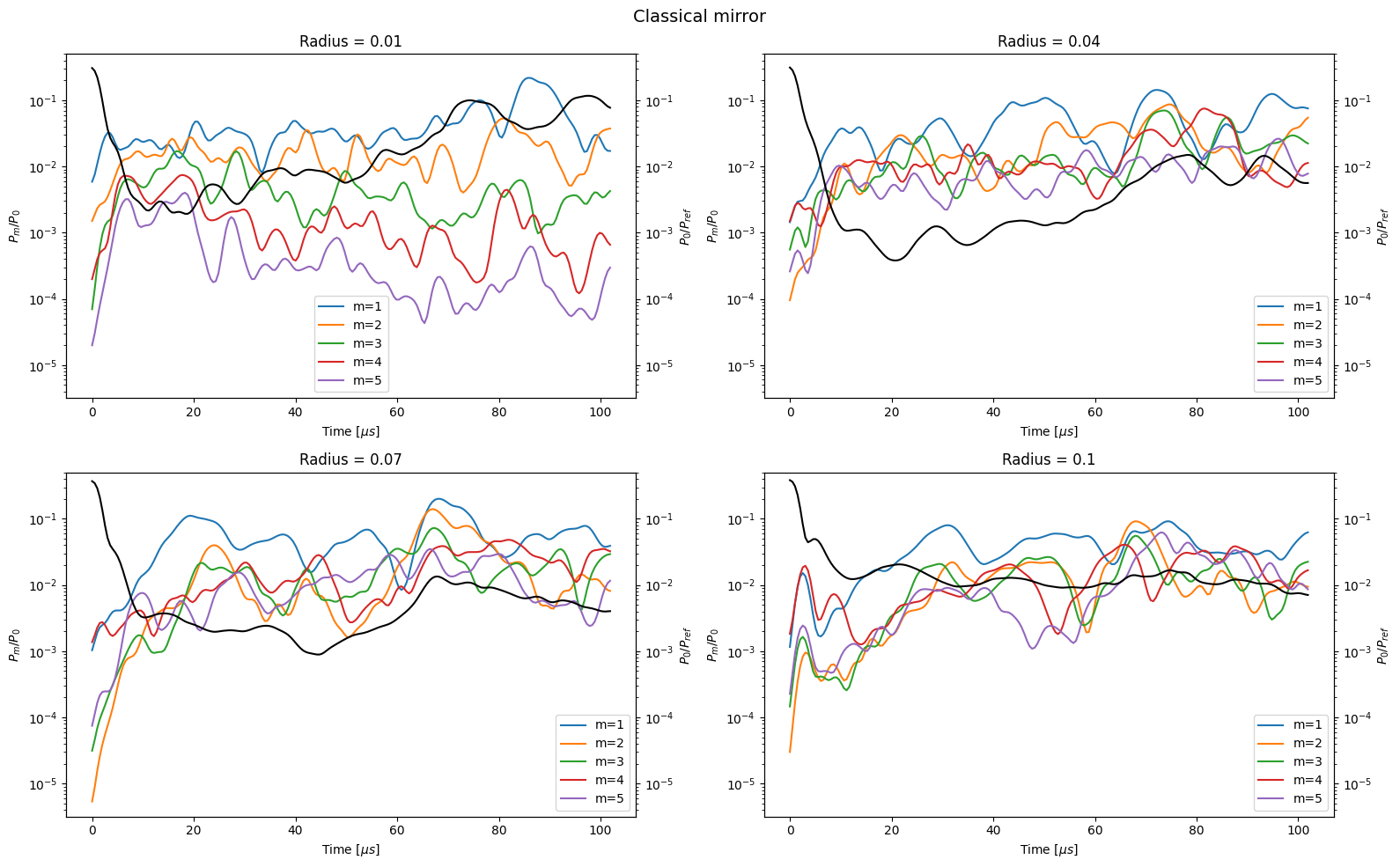}
    \caption{Classical mirror: power ratio $P_m/P_0$ with a Gaussian filter (standard deviation $\sigma=2$) of the azimuthal modes at specific radii. The black lines (right axes) show the time-series of the normalized zeroth azimuthal mode power $\hat{P} = P_0 / P_{ref}$ at specific radii, where $P_{ref} = P_0(t=5\mu s)$.}
    \label{fig:AzimuthalsClassicalTest2}
\end{figure}

\begin{figure}[H]
    \centering
    \includegraphics[width=0.9\textwidth]{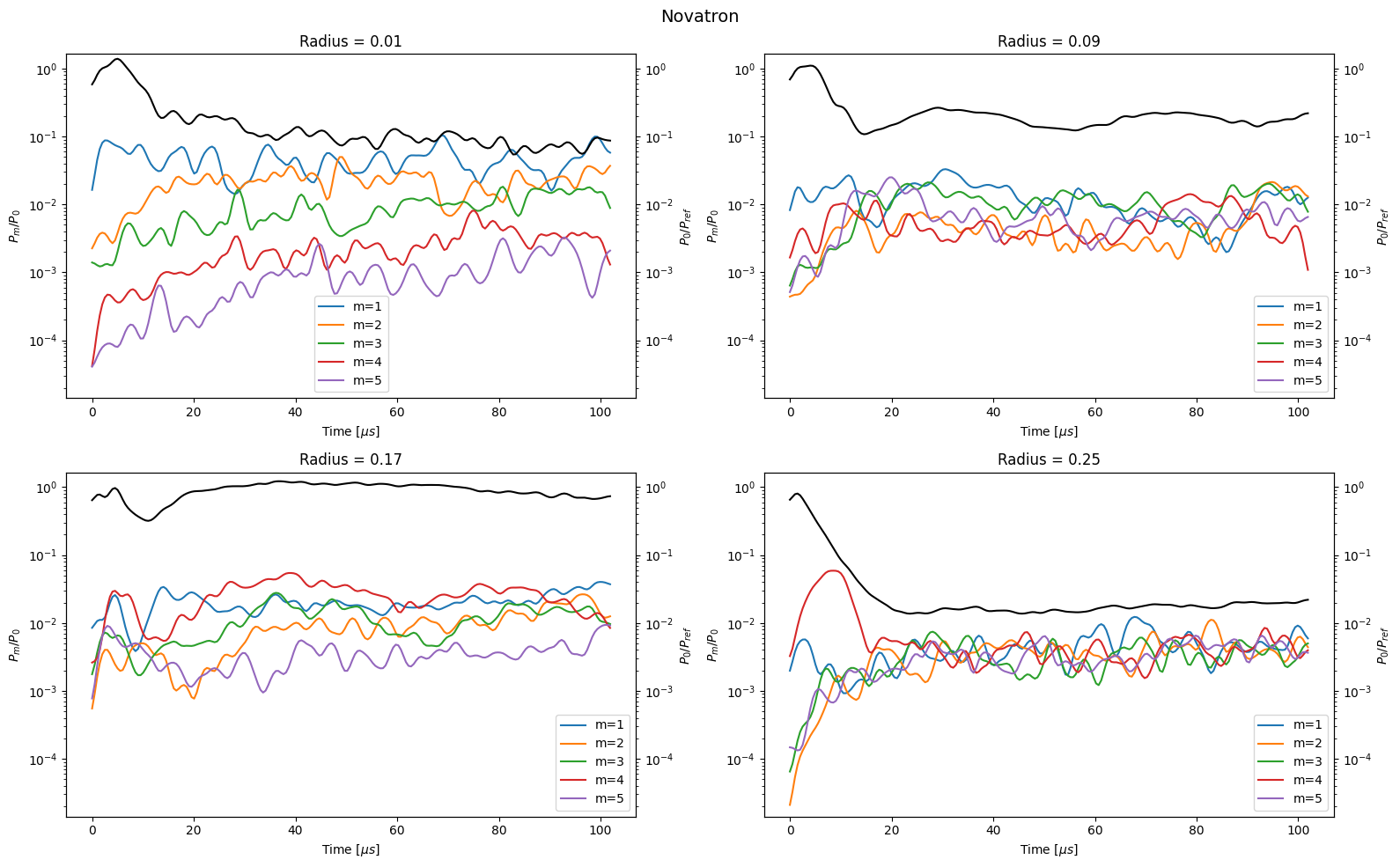}
    \caption{Same as Fig. \ref{fig:AzimuthalsClassicalTest2} but for the Novatron.}
    \label{fig:AzimuthalsN1Test2}
\end{figure}

The main difference in the stability analysis when comparing the Novatron simulations to the classical mirror is that for the Novatron no dominant global azimuthal mode has been established throughout the simulations, in contrast to what can be seen for the classical mirror, where $m=1$ dominates on all radii. To qualify this statement, the amplitudes of the azimuthal modes att different radii can be seen in Figures \ref{fig:AzimuthalsClassicalTest2} and \ref{fig:AzimuthalsN1Test2}, for the classical mirror and Novatron, respectively. In Figure \ref{fig:AzimuthalsClassicalTest2} the classical mirror $m=1$ mode power ratio increases to roughly $10^{-1}$ at all specified radii. In contrast, in the Novatron the largest $m=1$ mode power ratio is located at $r=0.01$ m, whilst at $r=0.17$ m the $m=4$ is largest, although comparable to the other modes. Throughout the Novatron simulation most modes oscillate around a power ratio around $10^{-2}$.

\begin{figure}[H]
    \centering
    \includegraphics[width=0.9\textwidth]{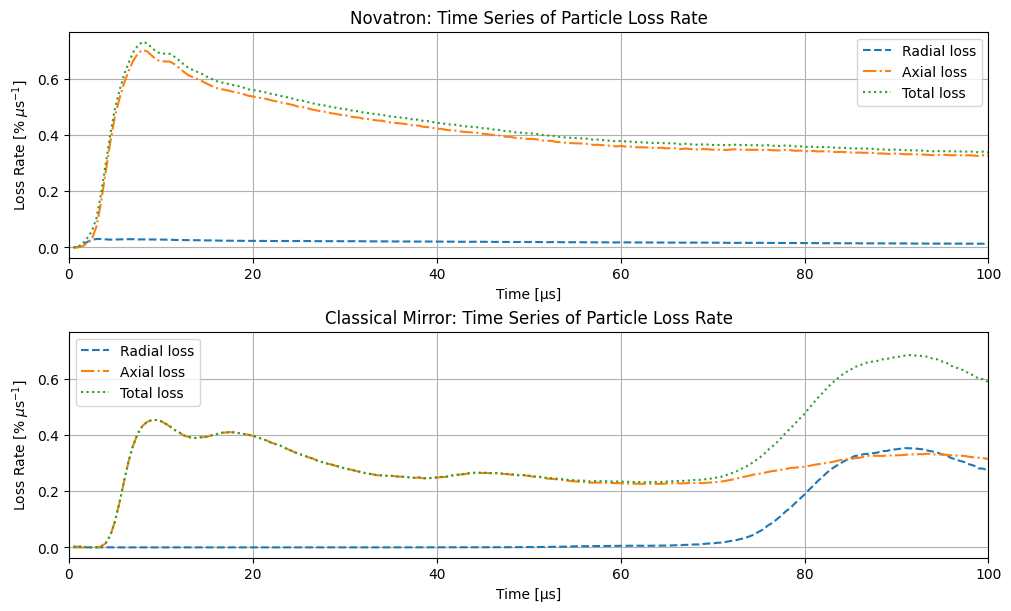}
    \caption{Loss rate [\% $\mu$$\text{s}^{-1}$] of particles, defined as $100( N_{\text{escaped}} / N_{\text{total}}) / dt$, in the Novatron (upper) and classical mirror (lower).}
\label{fig:loss_rate_test_2}
\end{figure}

Again, the only significant difference between the simulations of both machines is the external magnetic field. In the Novatron, in contrast to the classical mirror, we see a slow mixing of the plasma as it drifts azimuthally around the axis, similar to the effect of water circling a drain. This is most evident by the existence of the $m=1,2$ modes in the center, see Figure \ref{fig:AzimuthalsN1Test2}. These low azimuthal modes increase as plasma escapes in the non-adiabatic region, making the pressure profile more hollow near the magnetic null, and thus more unstable to interchange modes proximal to the axis. The increase in plasma loss from the adiabatic region contributes to the axial loss and is thus visible in the first 20 $\mu s$ or so in Figure \ref{fig:loss_rate_test_2}. As the azimuthal modes transport plasma inwards the pressure gradients become flattened (compare first to mid plot in Figure \ref{fig:RhoHybridWarpXNovatronClassicalTest2}), and thus naturally damp the interchange modes. The plasma will then reach an equilibrium between radial drift inwards and axial loss. Conversely, there is minimal radial drift outward towards the walls, as is evident from the number density in Figure \ref{fig:RhoHybridWarpXNovatronClassicalTest2} and the suppression of azimuthal modes in the outer region of the Novatron (see $r=0.25$ m in Figure \ref{fig:AzimuthalsN1Test2}).

In Figure \ref{fig:loss_rate_test_2} we see the loss rate of particles in both devices. Here, we see initially a better total ion confinement for the classical mirror, until the radial loss becomes more significant, so that the Novatron eventually outperforms in ion confinement. The higher loss rate in the Novatron is also explained by the fact that the plasma in regions of lower magnetic field is isotropic, thus a larger fraction of those ions fall into the loss cone. Figure \ref{fig:loss_rate_test_2} also shows that the Novatron, in line with previous cusp experiments, has a notable radial confinement of the plasma. This solidifies our prior belief that we should focus our efforts on decreasing the axial loss of the Novatron.

It should be noted that the maximum magnetic field strength in the classical mirror is $B=0.58$ T, and in the Novatron it is roughly $B=0.4$ T in the annular mirror. In the Novatron, the majority of the plasma will bounce between the annular mirrors. Thus, by running the same Novatron simulation with an increased magnetic field strength in this region to the same as the classical mirror will lead to a reduction in the maximal loss rate of roughly 24 \%. In Figure \ref{fig:loss_rate_test_2} we see the maximum axial loss rate at roughly 0.7 \%$\mu$s$^{-1}$ in the Novatron and 0.45 \%$\mu$s$^{-1}$  in the classical mirror. With the higher magnetic field value in the annular mirror the maximum loss rate is reduced to 0.53 \%$\mu$s$^{-1}$ . This gives us a maximum axial loss rate more comparable (although slightly higher) to the classical mirror.

\section{Discussion}

\subsection{Anisotropic equilibria and stability}
We have presented results for Novatron MHD stability employing anisotropic mirror equilibrium pressure profiles. The equilibrium equations have been discretized using a finite-difference method to properly resolve steep gradients at the plasma's edge.

The pressure profiles presented have the form $p(\psi, B) = A(\psi)p(B)$ for both parallel and perpendicular pressures. After a pressure profile solution has converged, it is analyzed with respect to the mirror, fire-hose, and interchange instability criteria. The mirror and fire-hose instabilities are automatically satisfied for all converged solutions, that is they set criteria for a well-posed system. The result is that the mirror and fire-hose instabilities impose a limit on the maximum $\beta_c$ value for each specific profile.

In terms of the interchange criteria, the pressure profiles have been evaluated using three flux line integral criteria termed Rosenbluth and Longmire, Generalized Rosenbluth and Longmire, and the CGL criterion. Based on these criteria, the studied pressure profiles in the Novatron magnetic field is found to be MHD stable throughout.

An important aspect not yet implemented in the equilibrium analysis is the ambipolar potential. This becomes important when the electrostatic energy is similar to the mean parallel ion energy, which is the case if a tandem-cell were to be added to the Novatron. The ambipolar potential can be included as a variable in the pressure profiles, for example $p(\psi,B,\phi)$. By introducing multiple species of ions and electrons that are trapped and lost, a potential can be computed, as was done in 1984 by Anderson and Rensink \cite{anderson_self-consistent_1984}.

Also, $\beta \sim 1$ conditions have not been included in this study. It has been shown in previous work by Fischer and Killeen (1971) \cite{fisher_finite_1971} that the level of anisotropy greatly limits the maximum obtainable $\beta$. In order to reach higher $\beta$ values, isotropic plasmas with pressure profiles $P(\psi)$ can be employed. Given its exceptional radial confinement, the Novatron is an ideal candidate for these $\beta$-enhancement studies.

\subsection{Hybrid-PIC simulations}
Two primary concerns with regards to the numerical methods underlying the hybrid-PIC code are the embedded boundary conditions and vacuum regions. The embedded boundary condition, as of writing, sets the electric field ad-hoc to zero on the boundary. This will result in an imperfect reflection of the electromagnetic waves. The problem is further amplified by the uniform grid which creates a ``stair-case'' effect at the embedded boundary. Two potential solutions to this problem are: 1) enhancing spatial resolution, particularly near the embedded boundary, and 2) creating regions of mirrored electromagnetic waves along and adjacent to the embedded boundary. The latter solution of mirrored waves are currently only implemented for the original 3D box boundaries, and not for the embedded boundary. Also, increasing the spatial resolution would require smaller time-steps and more macro-particles so as to have a smooth distribution function.

The inclusion of the embedded boundary, along with the static external magnetic field $B_v$, in the hybrid-PIC simulations has led to the appearance of artificial mode numbers in the electromagnetic fields. Notably, a locked $m=4$ mode is present in all simulations. This mode is most prominent at the beginning of the simulation and dampens over time as the plasma profiles relax. With the proper use of resistivity and hyper-resistivity, this artificial mode can be reduced to acceptable levels. Furthermore, the $m=4$ mode initially decreases until all modes are similar, and thus the embedded boundary in these simulations serve as an initial perturbation to the plasma profile.

To make definitive claims about the Novatron’s MHD stability, we compared each simulation with a roughly equivalent classical mirror simulation. This comparison isolates the effect of the external magnetic field, allowing us to evaluate the results against the theoretically known classical mirror MHD mode growth rates. It is important to note that improper boundary conditions can lead to pessimistic results for both machines.

Our WarpX hybrid-PIC simulations of the classical mirror and the Novatron yield two major findings. First, the classical mirror simulation confirms that the most deleterious mode $m=1$, becomes dominant with growth rates matching theoretical predictions. Second, the Novatron simulations show an absence of any dominant low MHD mode. However, there is a flattening of the plasma profile and fluctuations in the azimuthal modes as the plasma drifts inwards, or ``falls down" into the magnetic well. The main effect contributing to this mixing is the interchange instability, which occurs where the plasma pressure drops towards the axis, allowing plasma to spiral and drift inwards. This results in a balance between plasma escaping axially due to non-adiabatic effects and plasma mixing in the center.

\section{Conclusion}
Anisotropic magnetostatic equilibrium pressure profiles are, for the first time,  numerically computed for the Novatron mirror double-cusp configuration. These pressure profiles have been found to satisfy the fundamental mirror, fire-hose, and interchange stability criteria. Thus, the stability results presented here are consistent with, and expand upon, previous work on MHD stability of bi-conic cusps and classical mirrors. 

The plasma magnetic fields calculated from Eq. \eqref{eq:2} show an excavation (or well) produced in the total magnetic field. This effect will increase the mirror ratio of the machine, whilst also increasing the total magnetic field at the edges of the plasma, creating a steeper magnetic field gradient in the pedestal region. 

For each equilibrium pressure profile investigated, the maximum beta value at the mid-plane $\beta_c$ has been calculated. For the external magnetic field of the first generation Novatron, N1, employing the anisotropic pressure profiles Eqs. \eqref{eq:Novatron} we find $\beta_c=0.63$. This result corroborates previous numerical analysis of minimum-B mirrors in Ref. \cite{anderson_computation_1972}. However, this is not an experimental limit for the device, which can theoretically reach $\beta \sim 1$ conditions if the pressure profiles are isotropic, leading to $\beta$-enhancement effects at the plasma boundary \cite{hamamatsu_numerical_1983}.

The hybrid-PIC extension of the open-source WarpX code has been expanded to include a static external magnetic field and multi-cut polygon embedded boundaries. The anisotropic MHD equilibrium pressure profiles computed here, along with classical mirror equivalent profiles, were used as initial conditions for the hybrid-PIC simulations. 

The results from the classical mirror simulations confirm the analytical growth rates of the most deleterious $m=1$ azimuthal mode. For the Novatron, the results reveal an absence of any dominant low-MHD mode for a duration significantly exceeding the MHD interchange timescale. The plasma proximal to the axis undergoes mixing and azimuthal rotation, as evidenced by the fluctuations seen in the azimuthal mode analysis. On the outside of the plasma all MHD modes are damped, supporting the intuitive picture of strong stabilization in the favorable curvature vacuum magnetic field.

\section*{Acknowledgements}

This research used the open-source particle-in-cell code WarpX \url{https://github.com/ECP-WarpX/WarpX}, primarily funded by the US DOE Exascale Computing Project. Primary WarpX contributors are with LBNL, LLNL, CEA-LIDYL, SLAC, DESY, CERN, and TAE Technologies. We acknowledge all WarpX contributors.

\section*{Conflicts of Interest}

Novatron Fusion Group is a limited liability commercial company. All authors are fully or partially employed by Novatron Fusion Group and a sub-set owns shares or options in the company. The authors declare that they have no conflict of interest.

\bibliographystyle{iopart-num}
\section*{References}
\bibliography{references} 

\onecolumn 
\include{appendix}

\end{document}

%% file: appendix.tex
\section{Appendix}

In this Appendix, details of the derivation of the GRLI and CGLI interchange stability conditions are provided. 

\subsection{GRLI}

Inserting Eqs. (\ref{eq:p_v_delta}) into Eqs. (\ref{eq:deltaE1_2}) and (\ref{eq:deltaE2_2}) yields
\begin{align}
\Delta (dE_1) + \Delta (dE_2) &= p_1\Bigg(\frac{V_1}{V_1 + \delta V}\Bigg)^{\gamma} (V_1 + \delta V) \nonumber \\ 
&+ (p_1 + \delta p)\Bigg(\frac{V_1 + \delta V}{V_1}\Bigg)^{\gamma}V_1 \nonumber \\
&-p_1V_1 - (p_1 + \delta p)(V_1 + \delta V). \label{eq:triE}
\end{align}

\begin{figure}[h]
\centering
\includegraphics[width=78mm]{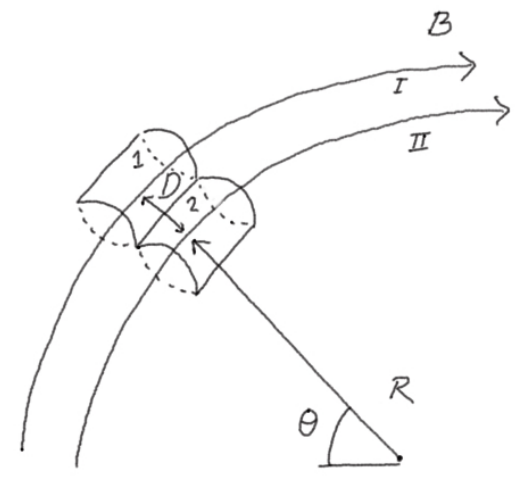}
\caption{Definition of the curvature radius $R$ and distance between tube centers $D$, see the text, drawn on top of two infinitesimal flux tubes 1 and 2. $B$ denotes the magnetic field.}\label{fig:R_D}
\end{figure}

We now set $p_1=p$,  $V_1=V$, and use the approximation

\begin{align}
(1 + x)^{\alpha} \approx 1 + \alpha x + \frac{\alpha (1 - \alpha)}{2}x^2 \text{ for small } x, \label{eq:Taylor_exp}
\end{align}

keeping terms up to 2nd order. We can thus write
\begin{align}
  \Delta (dE) & = \Delta (dE_1) + \Delta (dE_2) 
    = \int(\gamma-1)\delta V\Bigg(\delta p + \gamma p \frac{\delta V}{V}\Bigg) \label{eq:deltaEp}
\end{align} 
Next this expression needs to be simplified. We introduce the variables $D$, the distance between the centers of the infinitesimal tubes, or between the field lines on which the tubes are centered, and $R$, the curvature radius, see Fig. \ref{fig:R_D}. The direction of $R$ is taken to be the same for the two adjacent flux tubes. It follows Eq. (\ref{eq:phi}) that 

\begin{align}
    \delta V = \phi \delta\Bigg(\frac{d\ell}{B}\Bigg) = \phi\Bigg(\frac{\delta(d\ell)}{B} + d\ell\delta\Big(\frac{1}{B}\Big)\Bigg), \label{eq:delta_v}
\end{align}
\begin{align}
    \delta (d\ell) = (d\ell)_2 - (d\ell)_1 = 2\pi d\theta R - 2\pi d\theta(R + D) \\ 
    = -2\pi d\theta D = \Bigg\{d\theta=\frac{(d\ell)_1}{2\pi R}\Bigg\}  
    =-\frac{(d\ell)_1}{R}D. 
\end{align}

Setting $(d\ell)_1 = d\ell$ and using an identity originally derived in Rosenbluth 1957 \cite{rosenbluth_stability_1957}, which holds for $\nabla \times \textbf{B} = \textbf 0$,

\begin{align}
\frac{D}{R}=\frac{\delta B}{B},
\end{align}

gives 

\begin{align}
 \delta (d\ell) = -\frac{\delta B}{B}(d\ell).
\end{align}

Further, using Eq. (\ref{eq:phi}) and Eq. (\ref{eq:delta_v}),
\begin{align}
    \delta V = \phi\Bigg(-\frac{\delta B}{B^2}d\ell - d\ell \frac{\delta B}{B^2}\Bigg)  = -2\phi\frac{\delta B}{B^2}d\ell 
    = -2\phi \delta \Big( \frac{1}{B} \Big) d\ell, \label{eq:delV_delB_id}
\end{align}
\begin{align}
   \frac{\delta V}{V} = -\frac{2\phi\delta B d\ell}{B^2}\frac{B}{\phi d\ell} = -\frac{2\delta B}{B}.
\end{align}

With these expressions Eq. (\ref{eq:deltaEp}) can be simplified. The second factor in Eq. (\ref{eq:deltaEp}) can be written as:

\begin{align}
\delta p + \gamma p \frac{\delta V}{V} &= \delta p - 2p \gamma \frac{\delta B}{B} \\ \nonumber
&= B^{2\gamma} \Big( \frac{\delta p}{B^{2\gamma}} - p\frac{1}{B^{2\gamma + 1}} 2\gamma \delta B \Big) \nonumber \\
&= B^{2\gamma} \delta \Big( \frac{p}{B^{2\gamma}} \Big). 
\end{align}

Now, we can write

\begin{align}
   \Delta (dE) = \phi (\gamma -1) \delta\Bigg(\frac{1}{B}\Bigg)B^{2\gamma}\delta\Bigg(\frac{p}{B^{2\gamma}}\Bigg) d\ell
\end{align}

In order to arrive at a stability criterion, that is an expression for the total change in material energy associated with the flux tube interchange, $\Delta E$, we need to sum over all the infinitesimal flux cylinders along the tubes to require 

\begin{align}
	\Delta E = \Delta \int dE = \int \Delta (dE) = \int (\Delta (dE_1) +\Delta (dE_2))> 0  ,
\end{align}

The criterion can thus, omitting the constants $\phi$ and $(\gamma -1)$, be written as

\begin{align}
	\int \delta \Big(\frac {1}{B}\Big) B^{2 \gamma} \delta \Big( \frac {p_{\parallel} + p_{\perp}} {2B^{2 \gamma}}\Big) d\ell > 0.
\end{align}

\subsection{CGLI}

\subsubsection{From the CGL equations to expressions for $p_1'$ and $p_2'$}

Using Eqs. (\ref{eq:dad_1}) and (\ref{eq:dad_2}) and solving for $p_1'$ and $p_2'$ yields:
\begin{align}
    p_1' &= \frac{1}{2}(p_{\parallel, 1}' + p_{\perp,1}') \nonumber \\
    &=\frac{1}{2}\Bigg(\frac{p_{\parallel,1}B_1^2V_1^3}{B_2^2V_2^3} + \frac{p_{\perp 1}V_1B_2}{V_2B_1}\Bigg), \label{eqn:p_bis_1_simp}
\end{align}

\begin{align}
     p_2' &= \frac{1}{2}(p_{\parallel, 2}' + p_{\perp,2}'), \nonumber \\
    &=\frac{1}{2}\Bigg(\frac{p_{\parallel,2}B_2^2V_2^3}{B_1^2V_1^3} + \frac{p_{\perp,2}V_2B_1}{V_1B_2}\Bigg). \label{eqn:p_bis_2_simp}
\end{align}

\subsubsection{Rewriting $\Delta (dE)$ by expressing differences in the quantities between the flux tubes as perturbations}

Substituting Eqs. (\ref{eq:p_papa_delta}) - (\ref{eq:B_delta}) into Eq. (\ref{eq:delta_E_prim_only}) yields
\begin{align}
    \Delta (dE) &= \frac{1}{2}\Bigg[\frac{p_{\parallel 1}B_1^2V_1^3}{(B_1 + \delta B)^2(V_1 + \delta V)^2} + \frac{p_{\perp 1}V_1(B_1 + \delta B)}{B_1} \nonumber \\
    &- (p_{\parallel 1} + p_{\perp 1})V_1 +\frac{(p_{\parallel 1} + \delta p_{\parallel})(B_1 + \delta B)^2(V_1 + \delta V)^3}{B_1^2V_1^2} \nonumber \\
     &+ \frac{(p_{\perp 1} + \delta p_{\perp})(V_1 + \delta V)B_1}{(B_1 + \delta B)} - (p_{\parallel 1} + \delta p_{\parallel})(V_1 + \delta V) - (p_{\perp 1} + \delta p_{\perp})(V_1 + \delta V)\Bigg].
\end{align}

We set $p_{\parallel 1}=p_{\parallel}$,  $p_{\perp 1}=p_{\perp}
 $, $V_{ 1}=V$, $B_1 = B$, and drop the factor $1/2$. Gathering all $p_{\parallel}$ and $\delta p_{\parallel}$ terms, we define:

\begin{align}
 \Delta (dE_{\parallel}) &= \frac{p_{\parallel}B^2V^3}{(B + \delta B)^2(V + \delta V)^2} - p_{\parallel}V \nonumber + \frac{(p_{\parallel} + \delta p_{\parallel})(B + \delta B )^2 (V + \delta V)^3}{B^2 V^2} +
 (p_{\parallel 1} + \delta p_{\parallel})(V + \delta V).  
 \end{align}
 Next, using Eq. (\ref{eq:Taylor_exp}), neglecting terms of order 3 and higher, 
 we can write

\begin{align}
\Delta (dE_{\parallel}) &= 2\phi d\ell \delta B \frac{1}{B^2} \Bigg( - B^4 \delta \Big(\frac{p_{\parallel}}{B^4}\Big) \Bigg) \nonumber \\ 
&= - 2\phi d\ell \delta B B^2 \Bigg(\delta\Big(\frac{p_{\parallel}}{B^4}\Big) \Bigg) \nonumber \\
&= \Bigg\lbrace\delta \Big(\frac{1}{B} \Big) = -\frac{1}{B^2}\delta B \Bigg\rbrace \nonumber \\ 
&= 2\phi d\ell B^4 \delta \Big(\frac{1}{B} \Big) \delta \Big(\frac{p_{\parallel}}{B^4}\Big). 
\end{align}


Gathering all the $p_{\perp}$ and $\delta p_{\perp}$ terms, using a similar procedure as for the $p_{\parallel}$ terms we eventually get
\begin{align}
\Delta (dE_{\perp}) &= \phi d \ell \delta \Big(\frac{1}{B}\Big) B^3 \delta \Big(\frac{p_{\perp}}{B^3}\Big).
\end{align}

Thus, our new interchange stability criterion, derived from the CGL double adiabatic equations, becomes
\begin{align}
  \Delta E \propto \bigintsss\delta \Bigg(\frac{1}{B}\Bigg) \Bigg[B^4\delta\Bigg(\frac{p_{\parallel}}{B^4}\Bigg) + 
   \frac{B^3}{2}\delta\Bigg(\frac{p_{\perp}}{B^3}\Bigg)\Bigg]d\ell > 0. \label{eq:CGL_criterion}
\end{align}
where we have dropped the  $2 \phi$ factor. 


